\documentclass[12pt]{iopart}
\usepackage{graphicx} 
\usepackage{hyperref}
\hypersetup{
    colorlinks,
    linkcolor={red!50!black},
    citecolor={blue!50!black},
    urlcolor={blue!80!black}
}

\usepackage[bitstream-charter]{mathdesign}
\usepackage{xcolor}
\usepackage{bm}
\usepackage{physics}
\def\dd{{\rm d}}
\newcommand*{\tran}{^{\mkern-1.5mu\mathsf{T}}}  

\DeclareSymbolFont{usualmathcal}{OMS}{cmsy}{m}{n}
\DeclareSymbolFontAlphabet{\mathcal}{usualmathcal}

\begin{document}

\title[High-Dimensional Non-Convex Landscapes and Gradient Descent Dynamics]{High-Dimensional Non-Convex Landscapes and Gradient Descent Dynamics}

\author{Tony Bonnaire}
\address{Laboratoire de Physique de l’École normale supérieure, ENS, Université PSL,
CNRS, Sorbonne Université, Université Paris Cité, F-75005 Paris, France.}
\ead{tony.bonnaire@ens.fr}

\author{Davide Ghio}
\address{Ecole Polytechnique Fédérale de Lausanne (EPFL). IdePHICS Laboratory}
\ead{davide.ghio@epfl.ch}

\author{Kamesh Krishnamurthy}
\address{Joseph Henry Laboratories of Physics \& PNI,  Princeton University}
\ead{kameshk@princeton.edu}

\author{Francesca Mignacco}
\address{Joseph Henry Laboratories of Physics, Princeton University \& Initiative for Theoretical Sciences, Graduate Center, CUNY.}
\ead{fmignacco@princeton.edu}

\author{Atsushi Yamamura}
\address{Department of Applied Physics, Stanford University}
\ead{atsushi3@stanford.edu}

\author{Giulio Biroli}
\address{Laboratoire de Physique de l'Ecole Normale Sup\'erieure, ENS, Universit\'e PSL, CNRS, Sorbonne Universit\'e, Universit\'e Paris-Diderot, Sorbonne Paris Cit\'e, Paris, France}
\ead{giulio.biroli@ens.fr}

\vspace{10pt}
\begin{indented}

\item[]August 2017
\end{indented}
\section*{Abstract}
{\bf
In these lecture notes we present different methods and concepts developed in statistical physics to analyze gradient descent dynamics in high-dimensional non-convex landscapes. Our aim is to show how approaches developed in physics, mainly statistical physics of disordered systems, can be used to tackle open questions on high-dimensional dynamics in Machine Learning.   
}

\section{Introduction}  \label{sec:intro}
Analyzing high-dimensional dynamics is a central problem in modern science. It appears in very disparate fields: physics, biology, social science and machine learning. What is challenging is that many tools developed for studying the dynamics of few degrees of freedom do not transfer to the high-dimensional case. Several phenomena taking place in high-dimensional dynamics defy low-dimensional intuition. Developing a theory thus requires new paradigms, new methods and new concepts. 

Modern machine learning problems, the main context we focus on here, are at the center of this challenge. In fact, the number of data, their dimension, and the number of parameters used in machine learning algorithms are huge and are increasing in a steady manner over the years. Questions such as characterizing the loss landscape and the training dynamics are therefore central examples of the high-dimensional dynamics problem we have described above. 

In this set of notes we present methods and concepts developed in statistical physics, mainly statistical physics of disordered systems, which have proven to be useful in tackling high-dimensional dynamics. Needless to say, there is still much to do, to discover and to understand, which makes the research on these topics exciting and open to many new potential contributions. 

\section{A crash course in random matrix theory} \label{sec:RMT}
Large random matrices are ubiquitous in many domains ranging from physics to finance and biology \cite{Wigner1967, guhr1998random, Bouchaud2009}. They also play an important role in many modern problems of machine learning, mainly because of the large number of data and their high dimensionality \cite{couillet_liao_2022}. In this section, we shall recall some key properties of random matrices. We will then present a method -- the Dyson Brownian Motion -- which triggered substantial progress in recent years.

\subsection{The Gaussian orthogonal ensemble}

One of the most emblematic sets of random matrices is the Gaussian orthogonal ensemble (GOE). Matrices from this ensemble are symmetric and defined as
\begin{equation}
    \bm{M} = \frac{1}{\sqrt{2N}} (\bm{J} + \bm{J}\tran),
\end{equation}
with $j_{ij} \sim \mathcal{N}(0, \sigma^2)$. In other terms, the off-diagonal elements of $\bm{M}$, $m_{ij}$ with $i>j$ are independent Gaussian random variables with zero mean and variance $\sigma^2/N$. Diagonal elements $m_{ii}$, on the other hand, have twice this variance.
The probability distribution associated with the GOE can hence be written
\begin{align}
    p(\bm{M}) &= \frac{1}{Z} \prod_{i<j} \exp{-\frac{N}{\sigma^2} \frac{m_{ij}^2}{2}} \prod_i \exp{-\frac{N}{\sigma^2} \frac{m_{ii}^2}{4}},\\
    &= \frac{1}{Z} \exp{\sum_{i,j} -\frac{N}{\sigma^2} \frac{m_{ij}^2}{4}},\\
    &= \frac{1}{Z} \exp{-\frac{N}{4\sigma^2} \Tr \bm{M}^2}, \label{eq:GOE_measure}
\end{align}
where $Z$ is the normalization constant of the probability measure.
The measure defined by Eq.~\eqref{eq:GOE_measure} is invariant under orthogonal transformations of the matrix $\bm{M} \rightarrow \bm{O}\tran \bm{M} \bm{O}$ with $\bm{O} \in \mathbb{R}^{N\times N}$ an orthogonal matrix. This invariance explains the name \textit{orthogonal} given to the ensemble. In the following, we consider the case $\sigma=1$.

\subsection{Eigenvector and eigenvalue distributions of the GOE}

In many problems, one is actually interested in the spectrum of the matrix instead of its raw elements. Our first aim is hence to study the eigenvalues and eigenvectors of a matrix $\bm{M}$ drawn from the GOE. Both these quantities are random variables, and we denote $\bm{v}^{\alpha}$ and $\lambda_{\alpha}$ with $\alpha \in \{1, \cdots, N\}$ the eigenvectors and eigenvalues of $\bm{M}$ respectively. We also order the sets such that $\lambda_1$ is the largest eigenvalue and $\lambda_N$ is the smallest.

Let us discuss the probability distribution of eigenvectors first. The symmetry under orthogonal transformation established in Eq.~\eqref{eq:GOE_measure} implies rotational invariance. In consequence, the probability of any given eigenvector $\bm{v}^{\alpha}$ is uniform on the sphere of radius $\sqrt{N}$ and the elements of $\bm{v}^{\alpha}$ have variance $1/N$. By concentration of the Gaussian measure in high dimensions, the probability distribution on the sphere can be approximated by a Gaussian for each component $p(v^{\alpha}_i) \propto \exp{-N \left(v_{i}^{\alpha}\right)^2/2}$. This result can also be seen as a consequence of the the Gaussian annulus theorem\footnote{Stating that $p(\lVert v^{\alpha} \rVert_2 - \sqrt{N} \leq t) \geq 2 \exp{-ct^2}$.} or, physically, as the equivalence between micro-canonical and canonical measures.
The bottom line is that eigenvectors have a quite simple statistics, and do not present much interest for this ensemble of random matrices. 

We now focus on the eigenvalue distribution of $\bm{M}$. To do so, we first build a matrix stochastic process defined by
\begin{equation} \label{eq:process_GOE}
    \bm{M}(t + \dd t) = \frac{\bm{M}(t) + \bm{G}(t)}{\sqrt{1+\dd t}},
\end{equation}
where $\bm{G}(t)$ is a matrix from the GOE with $g_{ij} \sim \mathcal{N}(0, \dd t/N)$. From the independence of $\bm{M}(t)$ and $\bm{G}(t)$, the resulting matrix $\bm{M}(t+\dd t)$ is also belonging to the GOE and has, at any time $t$, the same variance as $\bm{M}(t)$ thanks to the normalization term.
Such a matrix stochastic process is called a \textit{Dyson-Brownian motion} (DBM) and its invariant measure gives the GOE by construction. Note also the different orders with respect to $\dd t$ in the terms of Eq.~\eqref{eq:process_GOE}; elements of the matrix $\bm{M}(t)$ are of order one while those of $\bm{G}(t)$ are of order $\sqrt{\dd t}$ by definition. Assuming $\dd t$ small, one can write
\begin{align}
    \bm{M}(t + \dd t) &\approx \frac{\bm{M}(t) + \bm{G}(t)}{1+\frac{\dd t}{2}}, \\
    &\approx \bm{M}(t) - \frac{1}{2} \bm{M}(t) \dd t + \bm{G}(t) + \mathcal{O}((\dd t)^{3/2}), \\
    &\approx \bm{M}(t) + \delta \bm{M}, \label{eq:perturbation}
\end{align}
where $\delta \bm{M} = -\frac{1}{2} \bm{M}(t) \dd t + \bm{G}(t)$ is a small perturbation of order $\dd t + (\dd t)^{1/2}$ of the matrix $\bm{M}(t)$. This can be equivalently written in terms of the eigenvalues of $\bm{M}(t)$, for all $\alpha \in \{1, \ldots, N\}$,
\begin{equation}
     \lambda_{\alpha}(t + \dd t) = \lambda_{\alpha}(t) + \delta \lambda_{\alpha},
\end{equation}
where $\delta \lambda_{\alpha}$ denotes the small perturbation associated to the eigenvalue $\lambda_{\alpha}$ of $\bm{M}(t)$. Using time-independent perturbation theory for matrices (or operators) and the notations from quantum mechanics, we have\footnote{This equation can be obtained by computing the eigenvalues of the power series development of Eq.~\eqref{eq:perturbation} for small perturbations and keeping only the two first terms.}
\begin{equation}
    \lambda_{\alpha}(t+\dd t) = \lambda_{\alpha}(t) + \bra{\alpha} \delta \bm{M} \ket{\alpha} + \sum_{\beta\neq \alpha} \frac{|\bra{\alpha}\delta \bm{M}  \ket{\beta}| ^2}{\lambda_{\alpha}(t) - \lambda_{\beta}(t)} + \mathcal{O}\left((\dd t)^{3/2}\right)\,,
\end{equation}
where we used the bra-ket notation for the eigenvectors $\bm{v}^{\alpha}$ of $\bm{M}(t)$. For instance, $\bra{\alpha} \delta\bm{M} \ket{\alpha}$ is the $\alpha \alpha$ element of $\delta\bm{M}$ in the basis diagonalizing $\bm{M}(t)$. The previous equation can be rewritten as
\begin{equation}
    \lambda_{\alpha}(t+\dd t) = \lambda_{\alpha}(t) - \frac{1}{2}\dd t \lambda_{\alpha}(t) + g_{\alpha\alpha} + \sum_{\beta\neq \alpha} \frac{g_{\alpha\beta}^2}{\lambda_{\alpha}(t) - \lambda_{\beta}(t)} + \mathcal{O}\left((\dd t)^{3/2}\right)\,,
\end{equation}
where, using the properties of the GOE, $g_{\alpha\alpha}$ and $g_{\alpha\beta}$ are uncorrelated gaussian variables (also in the basis of $\bm{M}$ because of rotational invariance). 
The previous equation is a discretized version of a stochastic equation. In order to consider its continuum limit, it is important to assess the order of magnitude of the different terms. The second term of the right-hand side is deterministic and of order $\dd t$, whereas the third term is Gaussian and of order $\sqrt{\dd t}$. These are indeed the usual scalings for stochastic equations. The fourth term can be written as 
\[
\sum_{\beta\neq \alpha} \frac{g_{\alpha\beta}^2}{\lambda_{\alpha}(t) - \lambda_{\beta}(t)}=\mathbb{E}_g\left[\sum_{\beta\neq \alpha} \frac{g_{\alpha\beta}^2}{\lambda_{\alpha}(t) - \lambda_{\beta}(t)}\right] + \text{fluctuations}.
\]
Since the fluctuations are of order $\dd t$, they can be neglected in the continuum limit (they give a sub-leading term with respect to $g_{\alpha\alpha}$). Therefore, one can replace the fourth term by its average $\frac{\dd t}{N}\sum_{\beta\neq \alpha} \frac{1}{\lambda_{\alpha}(t) - \lambda_{\beta}(t)}$. One therefore obtains the continuum limit stochastic equation on eigenvalues
\begin{equation}
    \frac{\dd \lambda_{\alpha}}{\dd t} = - \frac{\lambda_{\alpha}(t)}{2} + \frac{1}{N} \sum_{\beta\neq \alpha}^N \frac{1}{\lambda_{\alpha}(t) - \lambda_{\beta}(t)}+ \eta_{\alpha}(t),
\end{equation}
where $\eta_{\alpha}(t) $ is a white noise such that $\mathbb{E}[\eta_{\alpha}(t)\eta_{\beta}(t)] = 2 \delta_{\alpha,\beta}\delta(t-t')/N$.
In a more ``mathematically friendly'' way, we can write
\begin{equation}   \label{eq:langevin_eigs}
    \dd \lambda_{\alpha} = \left(- \frac{\lambda_{\alpha}(t)}{2} + \frac{1}{N} \sum_{\beta \neq \alpha}^N \frac{1}{\lambda_{\alpha}(t) - \lambda_{\beta}(t)} \right)\dd t + \dd B_\alpha \sqrt{\frac{2}{N}},
\end{equation}
where $\dd B_\alpha$ denotes the Brownian increment. Note this equation is valid for any $N$, although we will often consider the high-dimensional limit $N\to \infty$ in what follows. We also remark that Eq.~\eqref{eq:langevin_eigs} is a Langevin equation for $N$ interacting particles, that can be expressed as
\begin{equation} \label{eq:Langevin}
    \frac{\dd \lambda_{\alpha}}{\dd t} = - \frac{\partial V}{\partial \lambda_{\alpha}} + \eta_{\alpha}(t),
\end{equation}
with $\eta$ a Gaussian noise term with second moment
\begin{equation}
    \mathbb{E}\left(\eta_{\alpha}(t) \eta_{\alpha}(t^{\prime})\right) = 2 T \delta(t, t^{\prime}) \delta_{\alpha, \beta},
\end{equation}
where $T=1/N$ is the temperature. The potential $V$ is defined by
\begin{equation}
    -\frac{\partial V}{\partial \lambda_{\alpha}} = - \frac{\lambda_{\alpha}(t)}{2} + \frac{1}{N} \sum_{\beta \neq \alpha}^N \frac{1}{\lambda_{\alpha}(t) - \lambda_{\beta}(t)},
\end{equation}
leading to
\begin{equation} \label{eq:potential}
    V = \sum_{\alpha} \frac{\lambda_{\alpha}(t)^2}{4} - \frac{1}{N} \sum_{\alpha < \beta} \ln{|\lambda_{\alpha}(t) - \lambda_{\beta}(t)|}.
\end{equation}
We recognize two competing terms in $V$. The first one is a quadratic trapping potential forcing the eigenvalues to be close to zero. The second term is a repulsion term with the form of a two-dimensional Coulomb gas potential forcing the eigenvalues to space apart from each other. It implies in particular that all eigenvalues are correlated, making the system very complex to analyze with standard tools.

The stationary distribution of the process defined by Eq.~\eqref{eq:Langevin} is known to be given by the Boltzmann distribution
\begin{align}   
    p(\lambda_1, \cdots, \lambda_N) &=\frac{1}{Z} \exp{-\frac{V}{T}}, \\
    &=\frac{1}{Z}\exp{-N \sum_{\alpha} \frac{\left(\lambda_{\alpha}\right)^2}{4}} \prod_{\alpha < \beta} |\lambda_{\alpha} - \lambda_{\beta}|, \label{eq:joint_distrib_eigs}
\end{align}
hence describing the joint probability distribution of the eigenvalues of matrices belonging to the GOE, and where $Z$ provides the normalization constant.

\subsection{Density of eigenvalues}

The previous subsection tackled the problem of computing $p(\lambda_1, \cdots, \lambda_N)$, the joint probability distribution of eigenvalues. We may however wonder what is the typical density distribution of eigenvalues $\rho(\lambda)$ of a matrix taken from the GOE in the large $N$ limit.
Injecting the empirical measure
\begin{equation} \label{eq:DOE}
    \rho_N(\lambda) := \frac{1}{N} \sum_{\alpha} \delta(\lambda - \lambda_{\alpha})
\end{equation}
in the joint probability distribution, Eq.~\eqref{eq:joint_distrib_eigs} reads
\begin{multline} 
    p(\lambda_1, \cdots, \lambda_N) \propto \exp\biggl[\biggr.-N^2 \int \dd \lambda \frac{\lambda^2}{4} \rho_N(\lambda) 
    \\+ \frac{N^2}{2} \int \dd \lambda \dd \lambda^{\prime} \rho_N(\lambda) \rho_N(\lambda^{\prime}) \tilde{\ln} |\lambda - \lambda^{\prime}| - N \ln \epsilon \biggl.\biggr],
\end{multline}
where $\tilde{\ln}(|x|) = \ln \left(|x| + \epsilon \right)$\footnote{We have added this regularizer to take care of the term corresponding to $\lambda_i=\lambda_j$. At the end we will take the $\epsilon \rightarrow 0$ limit.  As we shall see, it will be possible to neglect the additional term $-N \ln \epsilon$ since it is subleading with respect to the ones in $N^2$.}.
The probability measure over the function $\rho_N(\lambda)$ can hence be written as a marginal distribution over the joint distribution of eigenvalues as
\begin{align}
    p(\rho_{N}(\lambda)) &= \int \prod_{\alpha=1}^N \dd \lambda_{\alpha} \delta[\rho_N(\lambda) - \rho(\lambda)] p(\lambda_1, \cdots, \lambda_N), \\
    &\propto \exp{-N^2 F(\rho_{N}(\lambda)) - N \ln \epsilon} \int \prod_{\alpha=1}^N \dd \lambda_{\alpha} \delta\left[\rho_N(\lambda) - \rho(\lambda) \right], \label{eq:prob_density_GOE}
\end{align}
where $\delta [ \cdot ]$ denotes the functional delta function, and $F$ reads
\begin{equation} \label{eq:functional_density}
    F(\rho_{N}(\lambda)) = \int \dd \lambda \frac{\lambda^2}{4} \rho_{N}(\lambda) - \frac{1}{2} \int \dd \lambda \dd \lambda^{\prime} \rho_{N}(\lambda) \rho_{N}(\lambda^{\prime}) \ln |\lambda - \lambda^{\prime} |.
\end{equation}
The first term of Eq.~\eqref{eq:prob_density_GOE} is an energetic contribution in which we can neglect the $N \ln \epsilon$ term in the large $N$ limit, as the leading contribution is of order $N^2$\footnote{This treatment of the $\epsilon$ term could look suspicious. The main point is that we are interested in the probability of the "macroscopic" density of eigenvalues, i.e. on scales of order one with respect to $N$.}. The second term is an entropic factor, that we will denote $S(\rho_N)$, accounting for the number of ways one can have $N$ particles giving a density $\rho_N(\lambda)$. It can also be seen as the Jacobian of the transformation going from an eigenvalue probability distribution to a functional of the density $\rho_N$.
To compute this entropic factor, let us first replace the delta distribution by its Fourier representation introducing an auxiliary (imaginary) function $g$ leading to
\begin{equation}
    S(\rho_{N}) \propto \int \prod_{\alpha=1}^N \dd \lambda_{\alpha} \int D g \exp{N \int \dd \lambda' g(\lambda') \rho_{N}(\lambda') - \sum_{\alpha=1}^N g(\lambda_{\alpha})},
\end{equation}
allowing to decouple the integral over all $\lambda_{\alpha}$'s. Indeed, one can now rewrite the exponential containing the $\lambda_{\alpha}$'s as a product and then perform independent integrals over the $\lambda_{\alpha}$'s:
\begin{align}
    S(\rho_N) &\propto \int Dg \exp{N \int \dd \lambda' g(\lambda') \rho_{N}(\lambda')} \int \prod_{a=1}^N \dd \lambda_{\alpha} \exp{-\sum_{\alpha=1}^N g(\lambda_{\alpha})},  \\
    &\propto \int Dg \exp{N \int \dd \lambda' g(\lambda') \rho_{N}(\lambda')} \left(\int \dd \lambda' \exp{-g(\lambda')} \right)^N, 
\end{align}
finally leading to
\begin{equation} \label{eq:entropy_GOE}
    S(\rho_{N}) \propto \int D g \exp{N \left[ \int \dd \lambda' g(\lambda') \rho_{N}(\lambda') + \log \int \dd \lambda' \exp{-g(\lambda')} \right]}.
\end{equation}
Performing a saddle-point on $S(\rho_N)$ requires to compute the functional derivative of the exponent in the exponential, which reads
\begin{equation}
    \int \dd \lambda' \rho_{N}(\lambda') \delta(\lambda-\lambda') - \frac{\int \dd \lambda' \delta(\lambda - \lambda') \exp{-g(\lambda')}}{\int \dd \lambda' \exp{-g(\lambda')}} = 0,
\end{equation}
hence leading to
\begin{equation}
    g(\lambda) = - \log \rho_{N}(\lambda) - \log Z,
\end{equation}
with $Z = \int \dd \lambda' \exp{-g(\lambda')}$.
Substituting it back into Eq.~\eqref{eq:entropy_GOE} finally gives
\begin{equation}
    S(\rho_{N}) \propto \exp{ -N \int \dd \lambda \rho_{N}(\lambda) \log \rho_{N}(\lambda)}.
\end{equation}
This entropic factor is consequently of order $N$ in the exponential meaning that it can be dropped out compared to the energetic contribution in $N^2$ when $N \rightarrow +\infty$. All in all, we have found that the probability density of $\rho_N$ is
\begin{equation}   \label{eq:large_deviation_rho_GOE}
    p(\rho_N(\lambda)) \propto \exp{-N^2 F(\rho_{N}(\lambda))}.
\end{equation}
This has the form of a large-deviation principle -- a natural thermodynamic result as $\rho_N$ is a macroscopic observable. It is interesting to remark that, contrary to the usual thermodynamic expression in which there is a $N$ in front of the intensive free-energy, we have got an $N^2$. The reason is that the temperature of the associated physical system is very small $T=1/N$.

\begin{figure}
    \centering
    \includegraphics[width=0.6\linewidth]{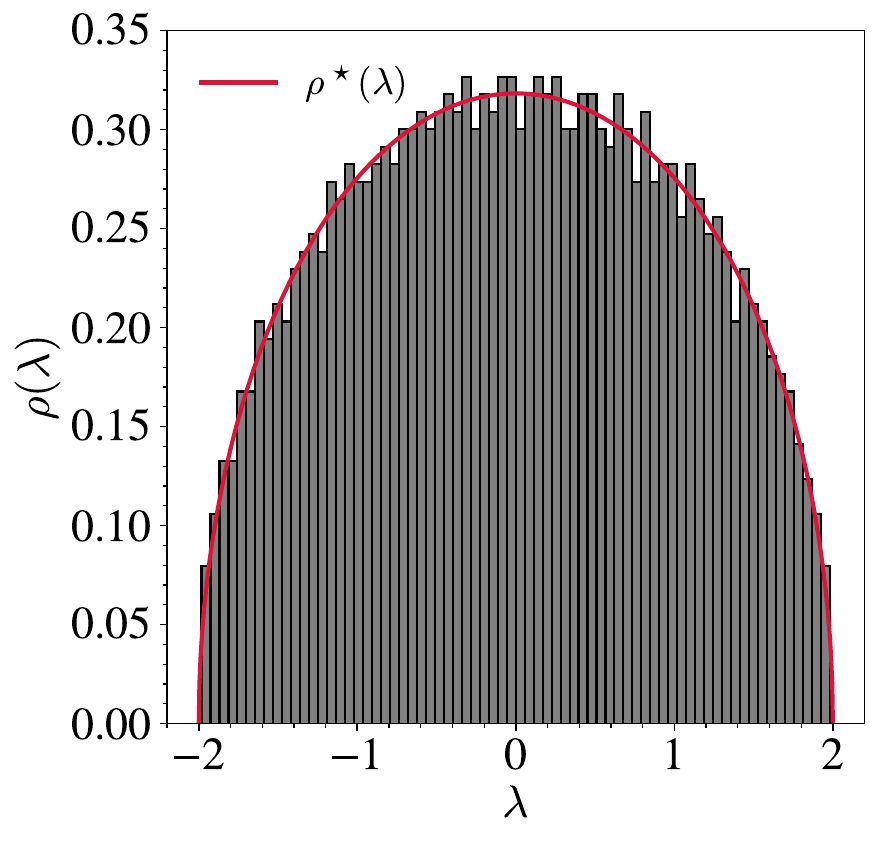}
    
    \caption{Density of eigenvalues for a matrix of the GOE. The solid red line displays the Wigner semi-circle law from Eq.~\eqref{eq:semi_circle_law} while the histogram shows the empirical distribution of eigenvalues obtained for a matrix with $N=2000$ and $\sigma^2 = 1$.}
    
    \label{fig:semi_circle}
\end{figure}

By concentration of measure arguments when $N \to \infty$, and a Laplace method on the set of plausible functions, we obtain that the density is a non-fluctuating quantity given by
\begin{equation}
    \rho_{N}(\lambda) = \rho^{\star}(\lambda) + \mathcal{O}\left(\frac{1}{N}\right),
\end{equation}
where the average $\rho^{\star}(\lambda) = \operatorname*{min}_{\rho(\lambda)} F(\rho(\lambda))$, under the constraint that $\rho^{\star}(\lambda)$ normalizes to one. To solve this variational problem, one needs to take the functional derivative of Eq.~\eqref{eq:functional_density} and set it to zero which yields
\begin{equation}
    \frac{\lambda^2}{4} - \int \dd \lambda' \rho^{\star}(\lambda') \ln |\lambda - \lambda^{\prime}| = 0.
\end{equation}
From there, one trick consists in taking the derivative of this latter expression with respect to $\lambda$ (taking care of the singularity in the logarithm term) giving
\begin{equation}
    \lim_{\Delta \to 0} \left[ \int_{-\infty}^{\lambda - \Delta} \dd \lambda' \frac{\rho^{\star}(\lambda')}{|\lambda - \lambda'|} + \int_{\lambda+\Delta}^{+\infty} \dd \lambda' \frac{\rho^{\star}(\lambda')}{|\lambda - \lambda'|}\right]= \frac{\lambda}{2}.
\end{equation}
The left-hand side of this equation is called the Cauchy principal value (noted $\mathrm{Pr}$) of the integral $\int \dd \lambda' \rho^{\star}(\lambda')/|\lambda - \lambda'|$. Therefore, the problem now boils down to finding $\rho^{\star}$ solution to
\begin{equation} \label{eq:Tricomi}
    \mathrm{Pr}\left[\int \dd \lambda' \frac{\rho^{\star}(\lambda')}{|\lambda - \lambda'|}\right] = \frac{\lambda}{2}.
\end{equation}
Actually, such an equation can be solved by Tricomi's theorem \cite{Tricomi1957}, see Ref. \cite{nadal2010phase} for the derivation.  The solution is 
\begin{equation}  \label{eq:semi_circle_law}
    \rho^{\star}(\lambda) = \frac{\sqrt{4 - \lambda^2}}{2 \pi}.
\end{equation}
For completeness, we also reinsert $\sigma$, whose dependence can be trivially deduced from the $\sigma=1$ result, to obtain the famous Wigner semi-circle law
\begin{equation}  \label{eq:semi_circle_law_sigma}
    \rho^{\star}(\lambda) = \frac{\sqrt{4\sigma^2 - \lambda^2}}{2 \pi \sigma^2},
\end{equation}
illustrated on Fig.~\ref{fig:semi_circle} as the solid red line together with an empirical distribution of eigenvalues obtained from an $N=2000$ and $\sigma^2 = 1$ simulation. 
Let us now comment on the finite but large $N$ case.  When $N$ is finite, the typical spacing of the eigenvalues in the bulk are of the order $1/N$, while they are of order $(1/N)^{2/3}$ at the edges of the spectrum located at $\pm 2\sigma$ \cite{guhr1998random}. We also see from Eq.~\eqref{eq:large_deviation_rho_GOE} that the fluctuations of the full distribution $\rho(\lambda)$ are scaling with $\exp{-N^2}$, meaning the convergence toward the global shape of the semi-circle law is very fast. However, for one eigenvalue to move away from the expected $\pm 2\sigma$ edge, Eq.~\eqref{eq:joint_distrib_eigs} teaches us that the cost is only exponential in $N$, as each eigenvalue contributes with a term of order $N$ to produce the overall $N^2$ contribution. 

So far, we have seen how to study random matrices belonging to the Gaussian orthogonal ensemble. Another interesting case, in particular for data sciences, is the Wishart ensemble with matrices of the form
\begin{equation}
    \bm{W} = \frac{1}{T} \sum_{\mu=1}^T \bm{\xi}_\mu \bm{\xi}_\mu\tran,
\end{equation}
where $\xi_{\mu,i} \overset{\mathrm{i.i.d.}}{\sim} \mathcal{N}(0,1)$. Such matrices are naturally arising as covariances of data sets but also as Hessian matrices of some single-layered neural networks. In this case, the eigenvalue density is known to follow the Marcenko-Pastur distribution \cite{Marcenko1967} when $N\rightarrow \infty$ and $T\rightarrow \infty$ simultaneously such that $q = N/T$ is finite, and
\begin{equation}
    \rho^{\star}(\lambda) = \frac{\sqrt{\left(\lambda_{+} - \lambda \right) \left( \lambda - \lambda_{-} \right) } }{2\pi q \lambda},
\end{equation}
where $\lambda_{\pm} = \left(1 \pm \sqrt{q} \right)^2$. The result above holds for $q<1$. For $q>1$ there is also a delta contribution located in the origin and with weight $1-1/q$ (the matrix $\bm{W}$ has rank $T$, and hence $N-T$ zero eigenvalues for $N>T$).
To obtain this result, and others, one can apply the previous procedure based on DBM and similarly build a stochastic process for which the joint probability of eigenvalues is a stationary state
\begin{equation}
    \xi_{\mu, i}(t + \dd t) = \frac{\xi_{\mu, i} + g_{\mu,i}}{\sqrt{1 + \dd t}}, 
\end{equation}
with $g_{\mu, i} \sim \mathcal{N}(0, \dd t)$.
There obviously exists other methods allowing the derivation of the quantities of interest from this section, for instance relying on the moment method, as done in the seminal paper by Wigner \cite{Wigner1955}, free probabilities \cite{Voiculescu1991} or on super-symmetry arguments \cite{Brezin1985}. However, the Dyson-Brownian motion method bridges well with the topic of this lecture on physics-inspired dynamics. It has also been shown particularly useful for the study of eigenvectors and how they evolve in other ensembles of random matrices, which also explains the recent resurgence of this particular approach in the literature \cite{erdHos2011universality}.

\subsection{Signal to noise ratio transition in random matrices}

We now apply DBM to derive a well-known signal-to-noise transition involving random matrices called the BBP phase transition \cite{baik2005phase} (the transition was first studied by the replica method in \cite{edwards1976eigenvalue}). We are interested in recovering a rank-one matrix (the signal) planted in a noisy background. The model is described as follows. Imagine that we are given a matrix $\tilde M$:
\begin{align}
    \tilde{M}_{i j}= M_{i j}+ \rho v_i v_j  \quad, \: |\mathbf{v}| = 1 \:\:,\: \bm{M} \sim \textrm{GOE}
\end{align}
Here, $\tilde{M}_{ij}, 1\geq i,j, \leq N$ are our measurements, $\mathbf{v}$ is the signal, $M_{ij}$ corresponds to the background noise, and the parameter $\rho$ controls the signal-to-noise. The matrix $\bm{M}$ is from the GOE, and its off-diagonal entries have a variance $1/N$. We are interested in the values of $\rho$ for which we can recover the signal $\mathbf{v}$, in the large $N$ limit.
This is a variant of the Principal Component Analysis problem, also called sometimes matrix PCA. 

To study this problem, we construct the DBM as follows: let $\tilde{\bm{M}}(t=0) = \rho \mathbf{v}\mathbf{v}^T$, and 
\begin{align}
    \tilde{\bm{M}}(t+dt) = \tilde{\bm{M}}(t) + \bm{g}(t)
\end{align}
where $\bm{g}(t)$ is a matrix from the GOE with a variance $dt/N$. By construction, $\tilde{\bm{M}}(t=1) = \tilde{\bm{M}}= \bm{M} +  \rho \mathbf{v}\mathbf{v}^T$, we have therefore to study this stochastic matrix process from $t=0$ to $t=1$. At $t=0$, there is only one non-zero eigenvalue $\lambda_1 = \rho$; at a a small but finite time, $\lambda_1 = \rho$ and the remaining eigenvalues are clustered around zero. As before, we can write a stochastic equation describing the evolution of the eigenvalues
\begin{align}
    \frac{d \lambda_\alpha}{d t}=\frac{1}{N} \sum_{\beta\neq \alpha} \frac{1}{\lambda_\alpha(t)-\lambda_\beta(t)}+\eta_\alpha(t).
\end{align}
For $\alpha > 1$, it can be rewritten as 
\begin{align}
    \frac{d \lambda_\alpha}{d t}=\frac{1}{N} \sum_{\beta\neq \alpha,1} \frac{1}{\lambda_\alpha(t)-\lambda_\beta(t)}
    +
    \frac{1}{N} \frac{1}{\lambda_\alpha(t)-\lambda_1(t)} +
    \eta_\alpha(t).
\end{align}
The first term is $O(1)$, the second term is $O(1/N)$ and the noise term is $O(1/\sqrt{N})$. In the large $N$ limit, we can thus ignore the second term, and the time evolution of $\lambda_{\alpha}$ for $\alpha>1$ is the same than for GOE: at time $t$ the $\{\lambda_{\alpha} \}$s will be distributed according to the semi-circle law with edges at $\pm 2\sqrt{t}$. Note that the reason the semi-circle has edges growing with $t$ is because we did not renormalize the variance after each infinitesimal step like before.


\begin{figure}
\begin{centering}
\includegraphics[scale=0.7
]{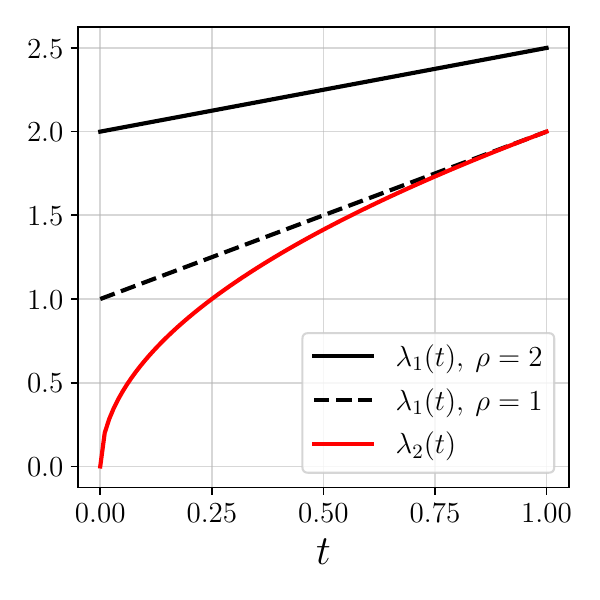}
\par\end{centering}
\caption{\label{fig:DBM_BBP} {\it Dyson-Brownian motion for the BBP transition.} Evolution of the outlier eigenvalue $\lambda_1$ (black lines) for two different values of $\rho$ and the edge of the semi-circle $\lambda_2$ (red line). When $\rho > 1$, the outlier is clearly separated from the bulk and the signal can be distinguished from the noise background. When $\rho=1$, there is a transition, and the outlier and the edge coincide for $t=1$. For $\rho<1$, there is no outlier, and the signal is lost in the noise background. 
}
\end{figure}


For the top eigenvalue, the equation reads
\begin{align}
    \frac{d \lambda_1}{d t}=\frac{1}{N} \sum_{\beta\neq 1} \frac{1}{\lambda_1(t)-\lambda_\beta(t)}
    +
    \eta_\alpha(t)
\end{align}
The first term is $O(1)$, and the noise term is $O(1/\sqrt{N})$. In the large $N$ limit we can make two more approximations: (i) neglect the noise term, and (ii) replace the sum in the first term with an integral over the distribution $\rho \left(\lambda_{\alpha>2} \right)$ -- the semi-cirle law with edges at $\pm 2\sqrt{t}$. The assumption (ii) holds as long as $\lambda_1$ is at a finite distance from all the other eigenvalues (finite means not vanishing when $N\rightarrow \infty$). 
With these approximations, we get
\begin{align}
    \frac{d \lambda_1}{d t}=\int d \lambda \: \rho(\lambda) \frac{1}{\lambda_1-\lambda}=\frac{\lambda_1(t)-\sqrt{\lambda_1^2(t)-2 t}}{2 t}.
\end{align}
The solution to this equation turns out to be simply given by
\begin{align}
    \lambda_1(t)=\rho+\frac{t}{\rho}.
\end{align}

It is important to ensure that the solution is consistent with the assumptions we made previously. In particular, we relied on a gap existing between $\lambda_1$ and the rest of the eigenvalues, i.e. $\lambda_1 - \lambda_2 > \epsilon > 0$, where $\lambda_2(t) = 2\sqrt{t}$ is the right edge of the semi-circle. Fig. \ref{fig:DBM_BBP} shows the behavior of the two eigenvalues for two different values of $\rho$. 

As explained before, we need to focus on $t=1$ to study the statistics of $\tilde M$. We find that for $\rho>1$ the signal is strong and the matrix $\tilde{\bm{M}}$ has one eigenvalue (a spike associated to the original signal) out of the Wigner semi-circle. For
$\rho=1$, the two largest eigenvalues meet at $t=1$ and we have a transition; for $\rho < 1$ reaches the edge of the semi-circle before $t=1$. In this case, one can show that for larger times $\lambda_1$ remains at the right edge of the semi-circle and the eigenvalue corresponding to the signal is buried within the bulk. 

Actually, to recover information on the signal, one has to focus on the eigenvector $\mathbf{v}_1$ corresponding to the largest eigenvalue. When $\rho > 1$, it can be shown that the leading eigenvector has a finite overlap with the signal direction $\mathbf{v}$ \cite{baik2005phase}. Specifically, 
\begin{align}
    \mathbf{v}_1 = \left( \sqrt{1 - \frac{1}{\rho^2}} \right) \mathbf{v} + \frac{1}{\rho} \mathbf{v}_{\perp}
\end{align}
where $\mathbf{v}_{\perp} $ is orthogonal to $\mathbf{v}$. However, when $\rho < 1$, the overlap of the leading eigenvector with the signal direction is vanishingly small : $\langle \mathbf{v}_1, \mathbf{v} \rangle \sim O(1/\sqrt{N})$. This can be understood studying the DBM for the eigenvectors (due to the small denominators in the perturbation theory the leading eigenvector hydridizes with all the other ones).   
Hence, $\rho=1$ corresponds to a transition from a regime in which the eigenvector associated to the largest eigenvalue has a finite component in the direction of the signal to a regime in which it does not.

\section{Gradient flow in matrix PCA and spherical spin-glasses} \label{sec:matrix_PCA}

We now use the signal+noise problem exposed in the previous section as a model to study how the dynamics of optimization in high-dimensions can lead to surprising phenomena. To do this, we formulate the problem of recovering the signal matrix in the noise background as an optimization problem and use gradient flow to find the optimum. Note that the purpose of doing this is to gain an insight on the dynamics and it does not imply that this is an efficient method to find the solution. We shall see that even in this simple problem one can gain interesting insights. In particular, we shall show that gradient flow has an algorithmic transition. Close to the transition, the system first converges toward uninformative saddles and then escape via a direction correlated with the signal. This is the simplest example of a more general mechanism that we will discuss later.  

Let us change slightly the notation of the signal+noise problem: our measurements $M_{ij}$ are given by 
\begin{align}
    M_{i j}=v_i v_j+\frac{N}{\rho} J_{i j},
\end{align}
where $\mathbf{v}$ is a vector such that $\| \mathbf{v} \|^2 = N$, $\rho$ is the signal-to-noise ratio and $\bm{J}$, which corresponds to the noise, is from the GOE. We aim at recovering $\mathbf{v}$ using gradient flow on the sphere. To do this we construct an energy function,
\begin{align}
    E(\mathbf{x}) = \frac{\rho}{4 N} \sum_{l m}\left(M_{l m}-x_l x_m\right)^2  \: ;
    \quad \sum_i x_i^2 = N,
\end{align}
and perform gradient flow on the sphere with random initial conditions. The $x$-dependent term of the energy function can be rewritten as 
\begin{align}
 &-\frac{\rho}{2 N} \sum_{l m} M_{l m} x_l x_m, \\
&= \underbrace{-\frac{1}{2} \sum_{l m} x_l J_{l m} x_m}_{\text {spin-glass term }} \: \: - \quad \underbrace{\frac{\rho}{2} N\left(\sum_l \frac{x_l v_l}{N}\right)^2}_{\text {deterministic term }}.
\end{align}
The first term has the form of an energy of a spherical spin glass.
The presence of the spin-glass term makes the landscape non-convex and leads to non-trivial dynamics (although a simple one due to the spherical constraint). 

We begin by analyzing the critical points of this energy function on the sphere. To do so, we form the Lagrangian given by
\begin{align}
\mathcal{L}\left( \mathbf{x}, \lambda \right)=E(\mathbf{x})+\frac{\lambda}{2}\left(\sum_i x_i^2-N\right).
\end{align}
where $\lambda$ is a Lagrange multiplier enforcing the spherical constraint. 
The extrema of the Lagrangian on the sphere satisfy
\begin{align}
    \frac{\partial E}{\partial x_i}+\lambda x_i=0 \quad ; \quad \sum_i x_i^2=N.
\end{align}
Let us define $\tilde{\bm{M}}= \rho \bm{M}/N = \bm{J} + \rho v v^{\top}$. Then, $E = -1/2 \sum_{l,m} x_l \tilde{M}_{lm} x_m$ and the equation for the extremum reads
\begin{align}
    -\sum_m \tilde{M}_{l,m} x_m + \lambda x_l = 0.
\end{align}
This is nothing but an eigenvalue equation with the corresponding eigenvalues $\{ \lambda^{\alpha} \}$ and eigenvectors $\mathbf{v}^{\alpha}$ as solutions. Therefore, this energy landscape has $2N$ critical points $\mathbf{x}^{\alpha} = \pm \sqrt{N} \mathbf{v}^{\alpha}$, each couple associated to an eigenvector and eigenvalue. 

\begin{figure}
\begin{centering}
\includegraphics[scale=0.5
]{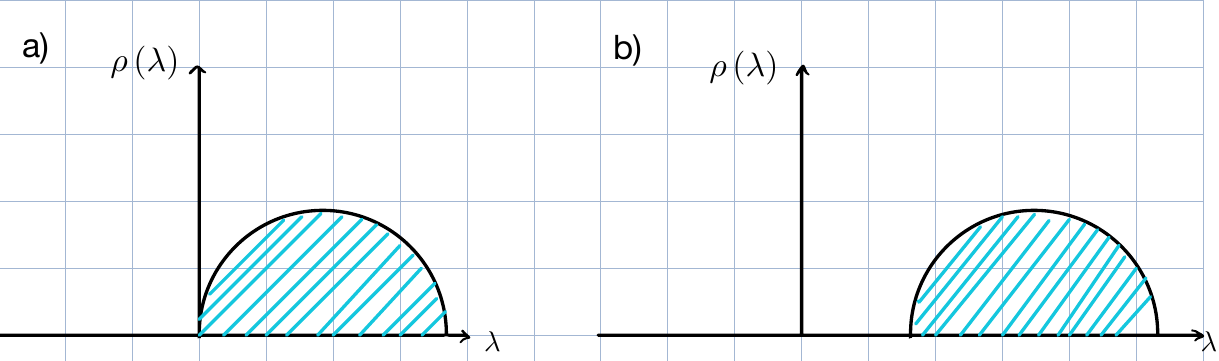}
\par\end{centering}
\caption{\label{fig:hess_spect} {\it Hessian spectra of the global minimum on the sphere.} a) when $\rho <1$, the global minima are marginally stable. b) when $\rho  > 1$, we have a stable global minimum
}
\end{figure}

The Hessian (evaluated on the sphere) of the critical point with index $\alpha$ is given by
\begin{align}
    H^{\alpha}_{lm} = -\tilde{M}_{lm} + \lambda^{\alpha} \delta_{lm},
\end{align}
and the eigenvalues of the Hessian are given by 
\begin{align}
    \epsilon^{\beta} = -\lambda^{\beta} + \lambda^{\alpha} \: ; \quad (\beta \neq \alpha).
\end{align}
Note there are only $N-1$ eigenvalues since the Hessian is calculated on the sphere. When $\alpha = 1$, all the eigenvalues $\epsilon^{\beta}$ are positive, and this corresponds to two global minima, whereas when $\alpha=N$, all the eigenvalues are negative, leading to two global maxima. All intermediate values of $\alpha$ have at least one unstable direction making them saddles. In this case, the landscape is hence simply composed of two maxima, two minima and $2N - 4$ saddles. We can now use the previous results obtained for the matrix $\tilde{\bm{M}}$ by the DBM method. 
Fig.~\ref{fig:hess_spect} shows the spectrum of the Hessian for the global minima when $\rho$ is less/greater than 1. When $\rho > 1$ we have a stable global minimum as there is a finite gap between the largest eigenvalue of $\tilde{\bm{M}}$ and the second largest (hence leading to $\epsilon^1>0$); this is not the case for $\rho<1$, in which there is no finite gap, $\epsilon^1 \rightarrow 0$ for $N\rightarrow 0$, hence leading to a marginally stable global minimum with many flat directions (or more precisely a Hessian characterized by arbitrary small positive eigenvalues). 
Moreover, when $\rho > 1$  the leading eigenvector $\mathbf{v}^1$, which points in the direction of the global minimum, has a finite overlap with the signal direction $\mathbf{v}$, whereas when $\rho < 1$, $\mathbf{v}^1$ is unaligned with $\mathbf{v}$.


The dynamics of the gradient flow on the sphere is given by
\begin{align}
    \frac{d x_i}{dt} = & \sum_m \tilde{M}_{i,m} x_m(t) - \lambda(t) x_i(t),
\end{align}
where the Lagrange multiplier $\lambda(t)$ enforces the spherical constraint on $\mathbf{x}$. Note that the Lagrange multiplier can be related to the energy as $\lambda(t) = -2E(t)/N$. We consider initial conditions sampled uniformly randomly on the sphere.  

Transforming to the eigenbasis of $\tilde{\bm{M}}$, we get 
\begin{equation}
    \frac{dx_{\alpha}}{dt} = \lambda_{\alpha}x_{\alpha}(t) - \lambda(t) x_{\alpha}(t),
\end{equation}
which implies
\begin{equation}
    x_{\alpha}(t) = x_{\alpha}(0) \exp{\int_0^{t} dt' (\lambda_{\alpha} - \lambda(t'))}.
\end{equation}

Moreover, the normalization constraint implies,
\begin{align}
N  = \sum_{\alpha} x_{\alpha}(0)^2 \exp{2\int_0^{t} dt' (\lambda_{\alpha} - \lambda(t'))},
\end{align}
which fixes $\lambda(t)$. To proceed, we first take the limit $N \to \infty$ (before $t \to \infty$). For large $N$, we can use the fact that the eigenvectors are uncorrelated with the eigenvalues, to write
\begin{align}
    1 = &  \sum_{\alpha} \frac{x_{\alpha}(0)^2}{N} \exp{2\int_0^{t} dt' (\lambda_{\alpha} - \lambda(t'))}, \\
     = &    \sum_{\alpha}  \exp{2\int_0^{t} dt' (\lambda_{\alpha} - \lambda(t'))}, \\
    = & \int d\eta \rho^*(\eta) \exp{2\int_0^{t} dt' (\eta - \lambda(t'))}, \\
= & \int d\eta \frac{\sqrt{2-\eta^2}}{2 \pi} \exp{2\int_0^{t} dt' (\eta - \lambda(t'))}, \\
     = &  \left( \int d\eta \frac{\sqrt{2-\eta^2}}{2 \pi} \exp{2\eta t}  
\right) \exp{-2\int_0^{t} dt'  \lambda(t')}.
\end{align}
This relation allows us to study the behavior of $\lambda(t)$ -- and thus the energy -- in different regimes. For instance, for large $t$, the term in the parenthesis is dominated by the largest exponent and to leading order it scales as $\exp{4t}/ t^{3/2}$, and this implies that for the RHS to be finite, we should have (for large $t$)
\begin{equation}
 \lambda(t) = 2 - \frac{3}{4t} + O\left( \frac{1}{t^2} \right),
\end{equation}
and hence
\begin{equation}
    E(t) = -1 + \frac{3}{8t} + O \left( \frac{1}{t^2} \right).
\end{equation}
We see that there is a power-law relaxation in the energy, and it reaches its lowest value of $-1$ slowly. How does this asymptotic value compare to the energy of the global minima $E_{gm}$? Using that 
$E_{gm}$ is given by $-1/2$ times the largest eigenvalue of $\tilde{\bm{M}}$, one finds that  
when $\rho > 1$, $E_{gm} = -1/2\,(\rho + 1/\rho) < -1$ and when $\rho < 1$, $E_{gm} = -1$. Thus, we find that for $\rho<1$ the system is able to reach the energy of the global minimum at large times. However, in this case the signal is not recovered as the global minimum does not point in the direction of the signal. On the contrary, for $\rho>1$ the global minimum does point in the direction of the signal but the system never reaches it in final time and, instead, remains at higher energies. 


\begin{figure}
\begin{centering}
\includegraphics[scale=0.65
]{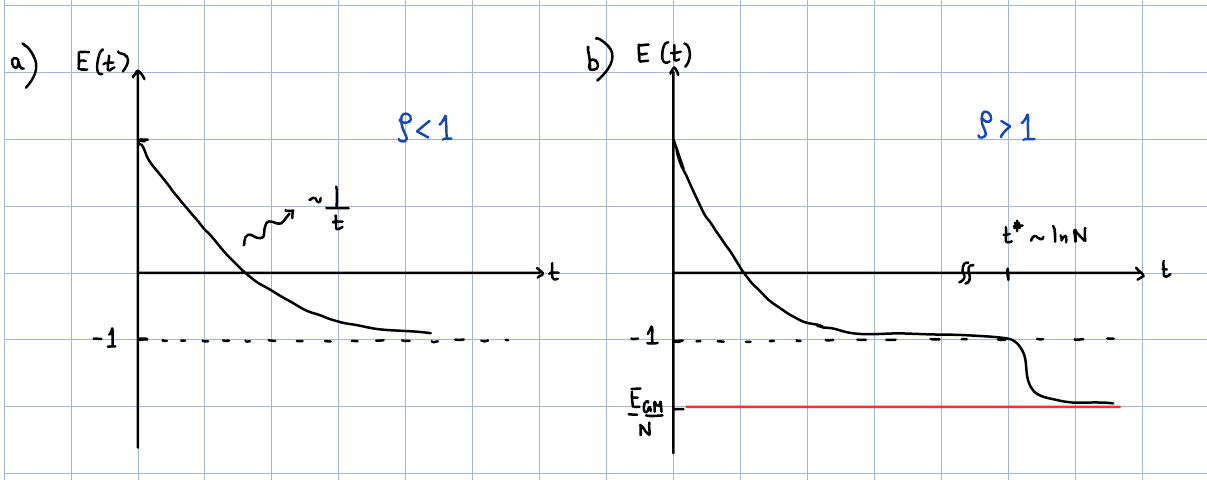}
\par\end{centering}
\caption{\label{fig:energy_dyn} {\it Energy dynamics:} (a) When $\rho < 1$ the energy relaxes to -1 as $\sim 1/t$ -- i.e. a power law. (b) When $\rho > 1$, there is a power law relaxation to -1, but after a time that grows as $t^* \sim \ln N$, the system will converge exponentially fast to the global minimum $E_{GM}/N < -1$.}
\end{figure}


In order to solve this puzzle, let us consider the behavior when $N$ is large, but not infinite. When, $\rho > 1$, the condition that enforces the spherical constraint is given by
\begin{align}
     1 = &  \left(  \sum_{\alpha \neq 1} \frac{x_{\alpha}(0)^2}{N}
     \exp{2\lambda_{\alpha}t}
     + \frac{x_{1}(0)^2}{N} \exp{2\lambda_{max}t}
     \right)  \exp{-2\int_0^{t} dt'  \lambda(t')}.
\end{align}
This gives us a timescale $t^*$ below which the first term dominates and, for $t \gg t^*$, the second term will dominate. The timescale is consequently given by
\begin{align}
    t^* \sim \frac{\ln N}{\lambda_{max} - 2}
\end{align}
Moreover, for $t \gg t^*$, $E(t) = -\lambda(t)/2 \approx -\lambda_{max}/2 $ meaning that the gradient flow eventually finds the global minimum, but only after a ``search phase'' which takes $\sim t^*$ time. This behaviour is sketched in Fig. \ref{fig:energy_dyn}. Such a behavior of the dynamics separated in two phases: a {\it search} phase (for $t<<t^*$) and a {\it convergence} phase (for $t>>t^*$) as it is also observed in more complex models and discussed in \cite{dAscoli2022a}.

In this section, we examined a quite simple model with a landscape that exhibits $2N$ critical points. The analysis of more complicated landscapes with exponential number of minimas is more involved and require more elaborated tools. This will be the topic of the next sections.

\section{Critical points of high-dimensional landscapes}
In many interesting problems, both in machine learning and in physics of disordered systems, one needs to deal with dynamics in various kinds of high-dimensional rough landscapes. Characterizing the properties of such landscape is generally a challenge. The analysis of mean-field spin-glasses provided a case in which such a challenge was faced for the first time \cite{cavagna1998stationary}. In recent years, this topic received a lot of interest from the mathematical and physics community. One of the main progress was the development of a method, called Kac-Rice, which allows to study in full details, and to a large extent rigorously, high-dimensional rough landscapes.  
\\ This section is divided as follows: in Sect. \ref{subsec:EnFunc} we define the kind of energy landscape we are going to study, motivating why it is important, and then in Sect. \ref{subsec:KacRice} is introduced the Kac-Rice method, which allows us to compute the complexity of critical points in such a framework. After going through the computation for a simplified case, we show the results for more general models, concluding with some references to applications of the Kac-Rice method beyond physics.
\subsection{Random Gaussian energy functions in high-dimensions and generalized spin-glasses}\label{subsec:EnFunc}
The energy function we are going to consider is of the form
\begin{equation} \label{eq:EnFunc_KR}
    E(\bm{s}) = -\sum_{p'=1}^{\infty} c_{p'}\sum_{i_1,\dots,i_{p'}} J_{i_1\dots i_{p'}}s_{i_1}\dots s_{i_{p'}} - rN\sum_{p'' = 1}^{\infty} \frac{b_{p''}}{p''} \left( \frac{1}{N}\sum_{i}s_iv_i\right)^{p''}\,,
\end{equation}
where $\bm{s}$ is a variable on the unitary sphere (i.e. $\sum_i s_i^2 = N$), $J_{i_1\dots i_{p}} \overset{\mathrm{i.i.d.}}{\sim} \mathcal{N}(0,\frac{1}{2N^{p-1}})$, $r$ plays the role of the Signal-to-Noise-Ratio and $\bm{v}$ is the signal that we want to retrieve. We notice that we have a first term which is random and a second term which instead is deterministic, and we will see how changing $r$, and thus changing the relative strength of the two, modifies the properties of the landscape. The energy function above can also be seen as the one of a generalized mean-field spin glass in the presence of a (generalized) field $\bm{\nu}$.
\\ In order to show the generality of this model, let us introduce two examples that can be described with this kind of formalism:
\begin{itemize}
    \item Tensor PCA \cite{richard2014statistical}: $p'=p''=p\geq3$. This is nothing else that the generalization to tensors of what we studied in the previous section, such that for example for $p=3$ one has $T_{i_1i_2i_3} = v_{i_1}v_{i_2}v_{i_3} + \lambda J_{i_1i_2i_3}$ and writing an energy function $E = \sum_{i_1,i_2,i_3}\left( T_{i_1i_2i_3} - s_{i_1}s_{i_2}s_{i_3} \right)^2$ it is easy to see that we come back to the form (\ref{eq:EnFunc_KR}) for $E(\bm{s})$, in which the parameter $\lambda$ acts as an inverse SNR (i.e. a Noise-to-Signal-Ratio).
    \item Random Gaussian Functions on the sphere (in high-dimension): Taking just one element of the first sum in (\ref{eq:EnFunc_KR}), we have $E(\bm{s}) = -\sum_{i_1,\dots,i_p}J_{i_1\dots i_p} s_{i_1}\dots s_{i_p}$, which is a Gaussian random variable on the sphere, since $\bm{J}$ is Gaussian and $\bm{s}$ is defined on the sphere. Therefore, we can easily compute 
    \begin{equation}
        \langle E(\bm{s}) \rangle_{\bm{J}} = 0\,,
    \end{equation}
    \begin{equation}
    \begin{aligned}
        \left\langle E(\bm{s})E(\bm{s'}) \right\rangle_{\bm{J}} &= \left\langle \sum_{i_1,\dots,i_p}J_{i_1\dots i_p} s_{i_1}\dots s_{i_p}\sum_{i'_1,\dots,i'_p}J_{i'_1\dots i'_p} s'_{i'_1}\dots s'_{i'_p}\right\rangle_{\bm{J}} = \\ &=  \frac{1}{2N^{p-1}} \sum_{i_1} s_{i_1}s'_{i_1} \dots \sum_{i_p} s_{i_p}s'_{i_p} =\frac{N}{2} \left( \frac{\sum_i s_i s'_i}{N}\right)^p = \frac{N}{2}q^p(\bm{s},\bm{s'})\,,
    \end{aligned}
    \end{equation}
    where we first used the independence of the elements of $\bm{J}$, then the value of their variance and finally we defined the \textit{overlap} $q(\bm{s},\bm{s'}) = \frac{1}{N}\sum_i s_i s'_i$. Thus, we have showed that the first term of (\ref{eq:EnFunc_KR}) is nothing else than the sum of Gaussian random functions of zero mean and covariance related to the overlap between two signals defined on the sphere. Putting back also the second (deterministic) term, which contributes only to the mean value, in general for (\ref{eq:EnFunc_KR}) we have
    \begin{equation}
        \langle E(\bm{s}) \rangle_{\bm{J}} = -rf_D(q(s,v))N\,,\quad \text{where}\quad f_D(x) = \sum_{p''=1}^{\infty} \frac{b_{p''}}{p''}x^{p''}\,,
    \end{equation}
    \begin{equation}
        \langle E(\bm{s}) E(\bm{s'}) \rangle_{\bm{J}} - \langle E(\bm{s}) \rangle_{\bm{J}} \langle E(\bm{s'}) \rangle_{\bm{J}} = \frac{N}{2}f_R(q(s,s'))\,,\quad \text{where}\quad f_R(x) = \sum_{p'=1}^{\infty} c_{p'}x^{p'}\,.
    \end{equation}
    Thus, the average of our Gaussian random function is going to depend only on the overlap between $\bm{s}$ and the signal $\bm{v}$ through a deterministic function $f_D$, which is defined in such a way that the overlap needs to be high in order to minimize the energy function. In the meantime, the second equation tells us that there is also a random part (of zero mean) with covariance depending just on the overlap between $\bm{s}$ and $\bm{s'}$.
\end{itemize}
We will now show a technique which allows us to study the structure of the critical points of this energy function and how it changes with $r$, $f_D$ and $f_R$.
\subsection{The Kac-Rice method}\label{subsec:KacRice}
The aim is to compute the number of critical points of $E(\bm{s})$, namely the number of vectors on the sphere $\bm{s'}$ such that $\nabla_{\perp}E(\bm{s'}) = 0$, where the symbol $\perp$ indicates that we are taking the gradient along the sphere.\\ We know that in one dimension the number of zeros of a certain function $f(x)$ can be written as $ \int \dd x\, \delta(f(x)) \,|f'(x)|$. The generalization in $N$ dimensions, for the zeros of the gradient, is then
\begin{equation}
    \mathcal{N}(E) \dd E = \int \dd\bm{s} \delta \left( \nabla_{\perp} E\right) |\det \nabla_{\perp}^2E(\bm{s})| \delta(E(\bm{s}) - E) \dd E\,,
\end{equation}
where we defined $\mathcal{N}(E)$ as the number of critical points with energy between $E$ and $E + \dd E$. Since $E(\bm{s})$ is a Gaussian random function, also $\mathcal{N}(E)$ is random, and in this section we are going to compute its average, following the pioneering work \cite{fyodorov2004complexity}. This average is  called \textit{annealed}, and we will write it as 
\begin{equation}
    \langle \mathcal{N}(E) \rangle \sim e^{N\Sigma (E/N)}\,,
\end{equation}
where $\Sigma$ is called \textit{complexity} (of critical points). In principle, one would be interested to go beyond just this average and use the fact that
\begin{equation}
    \mathcal{N}({E}) \sim e^{\langle \ln \mathcal{N}(E) \rangle}\,,
\end{equation}
but computing the \textit{quenched average} $\langle \ln \mathcal{N}(E) \rangle$ is in general a much more complicated task and can be done by combining the Kac-Rice and replica methods \cite{ros2019complexity}.
\subsubsection{Purely random case}
\begin{figure}
    \begin{minipage}{0.48\hsize}
        \begin{center}
    \includegraphics[width = \linewidth]{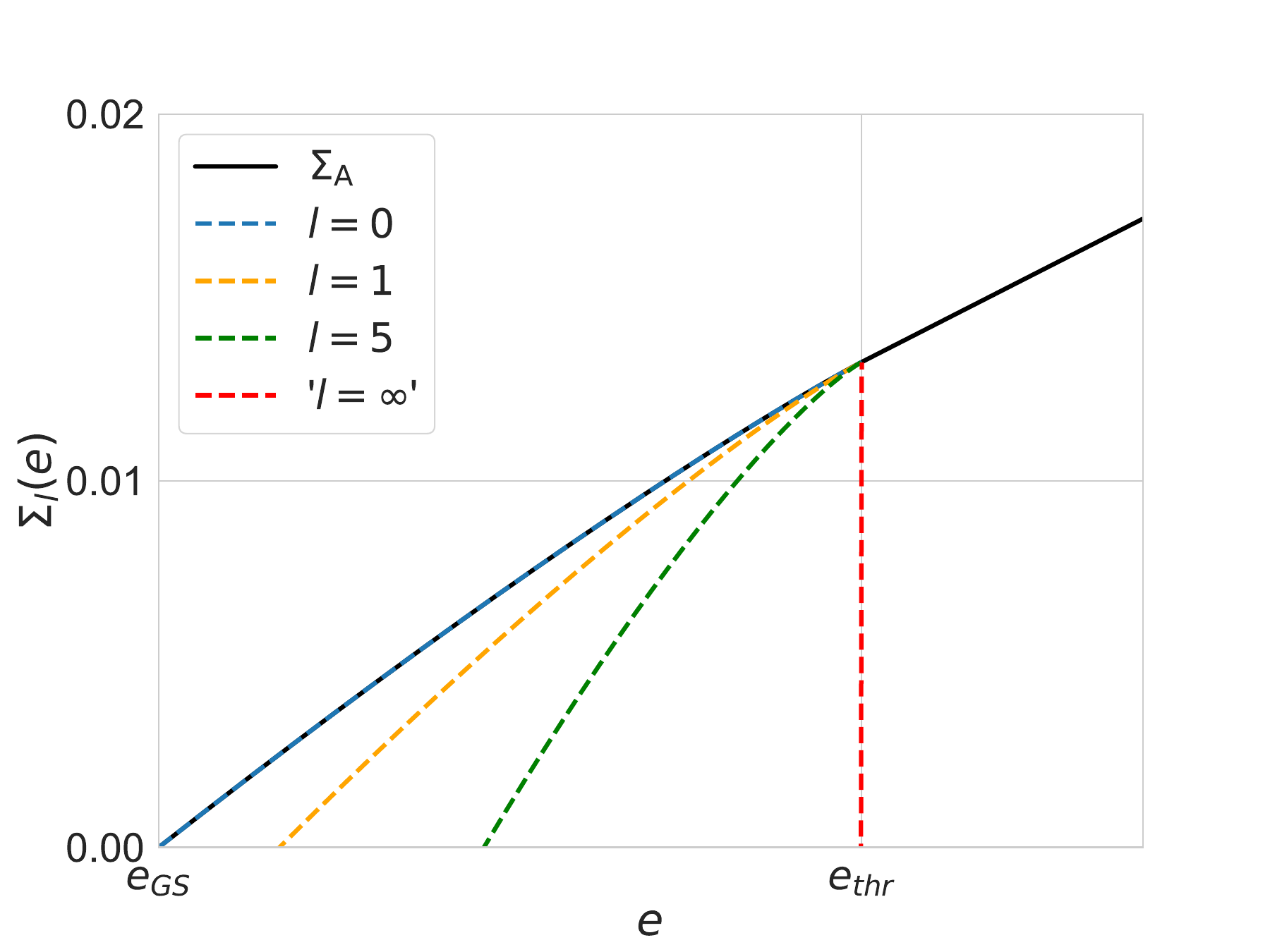}
        \end{center}
    \end{minipage}
    \begin{minipage}{0.48\hsize}
        \begin{center}
    \includegraphics[width = \linewidth]{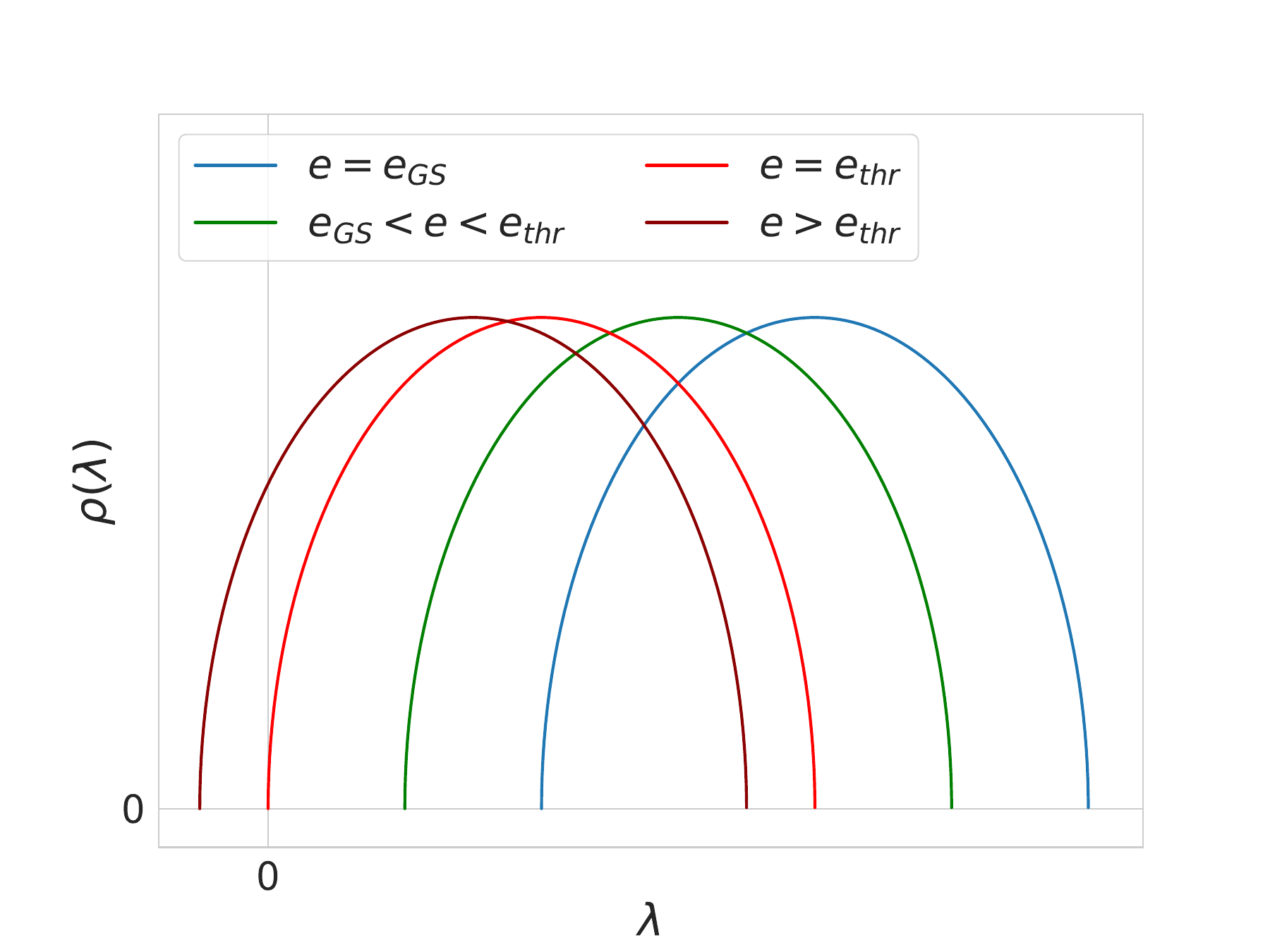}
        \end{center}
    \end{minipage}
    \caption{Left Panel: Behavior of the complexity of critical points for the spherical $p$-spin model with $p=3$. The plot was inspired from \cite{auffinger2020number}.\\Right Panel: Qualitative behaviour of the density of eigenvalues of the Hessians, given by shifted semi-circles, for different values of the energy $e$.}
    \label{fig:Comp_ps}
\end{figure}

Let us consider the case $f_D(x) = 0$, in which we can solve the problem in three, rather easy, steps. The computation we are going to show was first presented in \cite{fyodorov2004complexity}  .\\ First, we rewrite 
\begin{equation}\label{eq:Ps}
    \langle \mathcal{N}(E) \rangle = \int \dd \bm{s} P(\bm{s})\,, \quad \text{where} \quad P(\bm{s}) = \left\langle \delta \left( \nabla_{\perp} E\right) |\det \nabla_{\perp}^2E(\bm{s})| \delta(E(\bm{s}) - E) \right\rangle\,.
\end{equation}
Then, since we have $ \langle E(\bm{s}) \rangle_{\bm{J}} = 0 $ and $\langle E(\bm{s}) E(\bm{s'}) \rangle_{\bm{J}} = \frac{N}{2}f_R(q(s,s'))$, the probability in (\ref{eq:Ps}) depends just on the overlaps and as a consequence we have rotational invariance of $P(\bm{s})$, which thus needs to be uniform and we can write $P(\bm{s}) = P(\bm{1})\;\; \forall\,\bm{s}$. This means we can write $\langle \mathcal{N}(E)\rangle = P(\bm{1})\int \dd \bm{s} = P(\bm{1}) S_N(\sqrt{N})$, with $S_N(\sqrt{N})$ the surface of the $N$-dimensional sphere of radius $\sqrt{N}$, which can be computed in closed form and for $N$ large can be approximated as $S_N(\sqrt{N})\approx(2\pi e)^{N/2}$.\\ Thus, the second step is to compute 
\begin{equation} \label{eq:Pof1}
    P(\bm{1}) = \left\langle \delta \left( \nabla_{\perp} E(\bm{1})\right) |\det \nabla_{\perp}^2E(\bm{1})| \delta(E(\bm{1}) - E) \right\rangle\,.
\end{equation}
Considering again just one term of the first sum in (\ref{eq:EnFunc_KR}) by fixing $p'=p$, we see that \newline$E(\bm{1}) = -\sum_{i_1,\dots,i_p}J_{i_1\dots i_p}$ is clearly a Gaussian variable. But then $\nabla_{\perp} E(\bm{1})$ will be a Gaussian vector and $\nabla^2_{\perp} E(\bm{1})$ a Gaussian matrix, so that (\ref{eq:Pof1}) is nothing else than a Gaussian average of Gaussian variables, and thus we just need to compute their individual averages and covariances to solve (\ref{eq:Pof1}). First, since we neglected the deterministic part, we know that $\langle E(\bm{1}) \rangle = \langle \nabla_{\perp}E(\bm{1}) \rangle = \langle \nabla^2_{\perp}E(\bm{1}) \rangle = 0$ and we can also easily see that $\langle E(\bm{1})E(\bm{1}) \rangle = \frac{N}{2} f_R(1)$. Concerning the other covariances, we first use the fact that 
\begin{equation}
\langle\left( \nabla E(\bm{s})\right)E(\bm{s'}) \rangle = \nabla\langle E(\bm{s}) E(\bm{s'}) \rangle =\frac{N}{2}f'_R(q(s,s'))\frac{\bm{s'}}{N}
\end{equation}
to get $ \langle \left(\nabla E(\bm{1})\right)E(\bm{1})\rangle = \frac{f'_R(1)}{2} \bm{1}$. Then, 
in order to get $ \langle \left(\nabla_{\perp}E(\bm{1})\right)E(\bm{1})\rangle $ we project the LHS on the plane orthogonal to $\bm{1}$. However, repeating the same for the RHS, which is proportional to $\bm{1}$, we get zero hence establishing that $ \langle \left(\nabla_{\perp}E(\bm{1})\right)E(\bm{1})\rangle =0$. This means that that $\nabla_{\perp} E(\bm{1})$ is not correlated with $E(\bm{1})$. Similarly, one can obtain
\begin{equation}
    \langle \nabla^{\alpha}_{\perp} E(\bm{1})\nabla^{\beta}_{\perp} E(\bm{1})\rangle= \nabla^{\alpha}_{\perp}\nabla^{\beta}_{\perp} \langle E(\bm{1}) E(\bm{1})\rangle=\delta_{\alpha\beta}\frac{f'_R(1)}{2} + \mathcal{O}(1/N)
\end{equation}
and rewrite the Hessian as
\begin{equation}\label{eq:HessKR}
    [\nabla^{2}_{\perp} E(\bm{1})]_{\alpha\beta} = G_{\alpha\beta} - f'_R(1)\frac{E(\bm{1})}{N}\delta_{\alpha\beta}\,,
\end{equation}
where $G$ is a GOE matrix of covariance $\langle G^2_{\alpha\beta}\rangle = \frac{f''_R(1)}{2 N}$. From (\ref{eq:HessKR}) we see that the Hessian is correlated with the energy, and in particular one has
\begin{equation}
    \langle E(\bm{1}) [\nabla^{2}_{\perp} E(\bm{1})]_{\alpha\beta} \rangle = - f'_R(1)\frac{\langle E(\bm{1}) E(\bm{1}) \rangle}{N}\delta_{\alpha\beta} = - \frac{f_R(1)f'_R(1)}{2}\delta_{\alpha\beta}
\end{equation}
but not with the gradient, since it is easy to see that $\langle \nabla^{2}_{\perp} E(\bm{1})\nabla_{\perp} E(\bm{1}) \rangle = 0$.\\ Now we have all we need to compute $P(\bm{1})$. Defining $f_{\alpha}\equiv\nabla^{\alpha}_{\perp}E(\bm{1})$ and $\overline{e} \equiv E(\bm{1})/N$ we can write
\begin{equation}
\begin{aligned}
    P(\bm{1}) &= \int \prod_{\alpha=1}^{N-1} \dd f_{\alpha} \frac{e^{-f_{\alpha}^2/f'_R(1)}}{\sqrt{\pi f'_R(1)}} \dd \overline{e} \frac{e^{-N\overline{e}^2/f_R(1)}}{\sqrt{\pi f_R(1)/N}}\prod_{i,j}\dd G_{ij} \frac{e^{-G_{ij}^2/f''_R(1)}}{\sqrt{\pi f''_R(1)}}\prod_{\alpha}\delta(f_{\alpha}) \cdot \\ &\cdot |\det(G_{\alpha\beta} - f'_R(1)\overline{e}\delta_{\alpha\beta})|\delta(e-\overline{e}) = \\ &= \left( \frac{1}{\sqrt{\pi f'_R(1)}} \right)^{N-1} \frac{e^{-Ne^2/f_R(1)}}{\sqrt{\pi f_R(1)/N}} \langle|\det(G - f'_R(1)e\mathbb{I})|\rangle_{\text{GOE}}\,,
\end{aligned}
\end{equation}
where we used the delta distributions and we wrote the integral over the $G_{ij}$ implicitly.\\ The third and final step is to compute explicitly this GOE average, using the techniques we described in Section~\ref{sec:RMT}. This could be done in general for every $N$, but what we are going to do is to take the limit $N\rightarrow\infty$. Then, calling $\lambda_{\alpha}$ the eigenvalues of $G$ we can rewrite the average as 
\begin{equation}\label{eq:GOEint}
    \left\langle \prod_{\alpha} |\lambda_{\alpha} - f'_R(1)e|\right\rangle = \left\langle \exp\sum_{\alpha} \ln|\lambda_{\alpha} - f'_R(1)e|\right\rangle = \left\langle \exp \left(N\int \dd \lambda \rho(\lambda)\ln|\lambda - f'_R(1)e|\right)\right\rangle
\end{equation}
where we used the definition of the density of eigenvalues (\ref{eq:DOE}). Now in general this average is still complicated to compute, but in the large $N$ limit we can exploit the properties of GOEs that we reported in Section~\ref{sec:RMT}; namely that $\rho(\lambda)$ converges to the semicircle law $\rho^*(\lambda)$ as $e^{-N^2}$, which allows us to erase the average in (\ref{eq:GOEint}) just by substituting the semicircle law to the density $\rho(\lambda)$, since all the fluctuations become exponentially small for $N\rightarrow\infty$. Then writing explicitly $\rho^*(\lambda)$, we get
\begin{equation}
     \exp \left(N\int \dd \lambda \frac{\sqrt{2f''_R(1)-\lambda^2}}{\pi f''_R(1)}\ln|\lambda - f'_R(1)e|\right)\,.
\end{equation}
Putting everything together, we finally get
\begin{equation}
    \begin{aligned}
        \langle \mathcal{N}(E) \rangle = &\exp \Bigg( N\Bigg[ \left( \frac{1}{2} \ln \pi -\frac{1}{2}\ln\frac{1}{2} + \frac{1}{2}\right) -\frac{1}{2} \ln\pi f'_R(1) - \frac{e^2}{f_R(1)} + \\ &+ \int \dd \lambda \frac{\sqrt{2f''_R(1)-\lambda^2}}{\pi f''_R(1)}\ln|\lambda - f'_R(1)e|\Bigg] \Bigg) \equiv \exp\left(N\Sigma(e)\right) \,,
    \end{aligned}
\end{equation}
where we used that $S_N(\sqrt{N})=_{N>>1} (2\pi e)^{N/2}$ in the first step, the second and third terms come from the second step, and the last one from the third step.\\ We can now use this expression to see how the complexity behaves for some particular examples, starting with the case in which $f_D(x)=0$ and $f_R(x)=x^p$, such that $E(\bm{s}) = -\sum_{i_1,\dots,i_p}J_{i_1\dots i_p}s_{i_1}\dots s_{i_p}$, which is the well-known spherical $p$-spin model. For such a problem it was proven, first with arguments from physics in \cite{cavagna1998stationary} and then rigorously in \cite{auffinger2013random, subag2017complexity, subag2021concentration}, that the annealed and the quenched averages coincide; i.e. $\Sigma_A(e) = \Sigma_Q(e)$. In such a case one can also compute the complexities of specific kinds of critical points; namely separating them depending on their index value $l$, describing the number of negative directions along which the function increases. For example, $\Sigma_{l=0}$ is the complexity of the minima, while $\Sigma_{l=1}$ is the complexity of the saddle points with just one direction going down and all the others going up, and so on. What one finds, depicted in Fig.~\ref{fig:Comp_ps}, is that there exists a threshold $e_{thr}$ under which the minima dominate ($\Sigma_A=\Sigma_0$), and the eigen-spectrum of the Hessian is a shifted semi-circle that does not touch zero, and over which there are no minima with probability $1$ but only saddles with a finite fraction of directions which go down, and the Hessian's eigen-spectrum includes negative values. This implies that the energy landscape shows a sub-exponential number of minima for $e=e_{GS}$, then for $e_{GS}<e<e_{thr}$ an exponential number of minima appear, until at $e=e_{thr}$ they become marginally stable and after this point they disappear and the landscape becomes dominated by saddles with a finite fraction of directions going down.
\subsubsection{Results for the general case}
\begin{figure}[th!]
    \centering
    \includegraphics[width=0.8\linewidth]{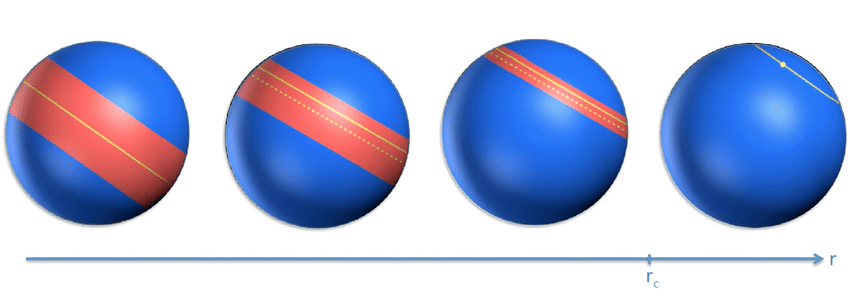}
    \caption{Behaviour of the energy landscape in the case in which $f'_D(0)>0$. This drawing, taken from \cite{ros2019complex}, illustrates the evolution of the energy landscape due to the increase of $r$. The red strip denotes the region on the sphere where minima lie in an exponential number. The continuous yellow line corresponds to the parallel where the deepest minima are located. The dashed yellow line corresponds to the parallel where the most numerous minima are located. At $r_C$, the energy landscape has a transition: For $r<r_C$, it is rough and full of minima; for $r>r_C$, it is smooth and contains only one minimum (represented by the yellow dot in the figure).}
    \label{fig:EL1}
\end{figure}

\begin{figure}
    \centering
    \includegraphics[width=0.8\linewidth]{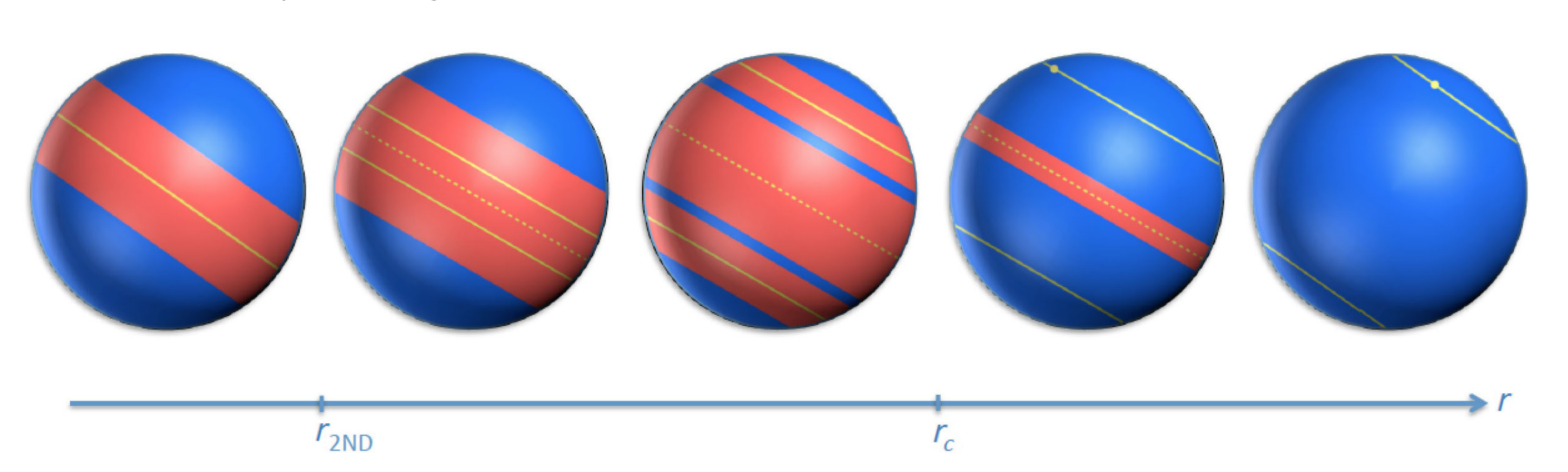}
    \caption{Behaviour of the energy landscape in the case in which $f'_D(0)=0$ and $f''_D(0)>0$. This drawing, taken from \cite{ros2019complex}, illustrates the evolution of the energy landscape due to the increase of $r$. The red strip denotes the region on the sphere where minima lie in an exponential number. The continuous (dashed) yellow line corresponds to the parallel where the deepest (most numerous) minima are located. The energy landscape has several transitions. At $r_{2ND}$, the deepest minima are no longer on the equator and move toward the poles. Afterwards, the band containing the exponential number of minima fractures into three parts, one around the equator and two symmetric ones closer to the poles. At $r_C$, the bands closer to the pole implode and are replaced by two isolated global minima (the one on the south hemisphere is not visible, since it is on the back of the sphere), but the band at the equator persists. Finally, for even larger values of $r$, the landscape becomes completely smooth with only two symmetric minima.}
    \label{fig:EL2}
\end{figure}

\begin{figure}[th!]
    \centering
    \includegraphics[width=0.8\linewidth]{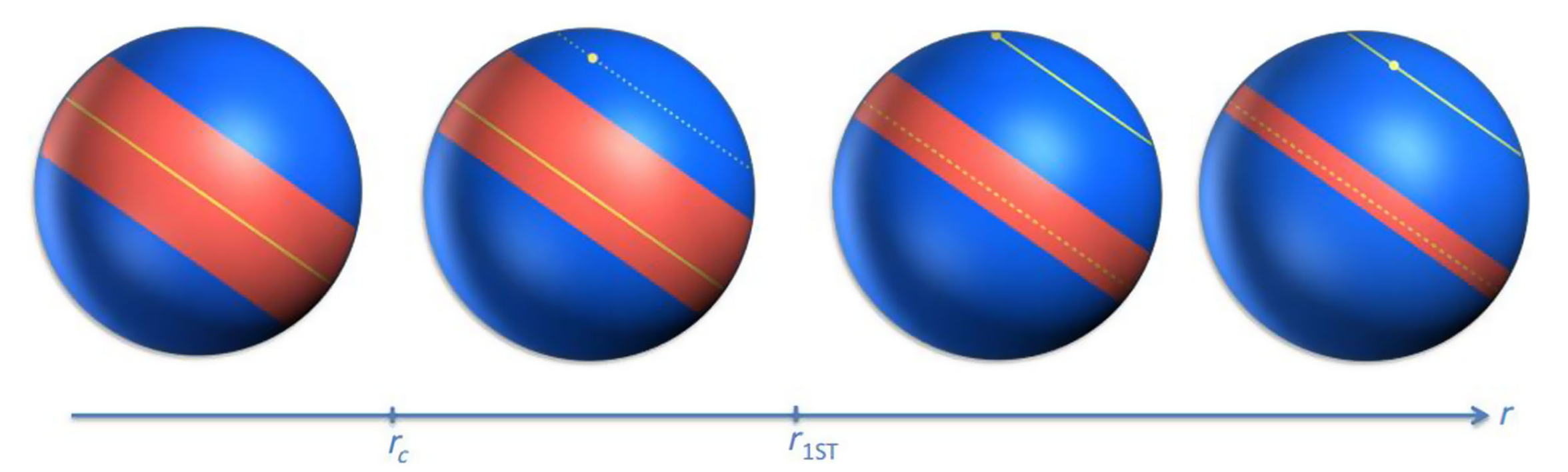}
    \caption{Behaviour of the energy landscape in the case in which $f_D(x)=x^3/3$ and $p=3$. This drawing, taken from \cite{ros2019complex}, illustrates the evolution of the energy landscape due to the increase of $r$. The red strips denote the regions on the sphere where minima lie in an exponential number. The continuous yellow line corresponds to the parallel where the global minimum is located. At $r_C$, an isolated local minimum appears. The dotted yellow line denotes that it is not yet the global one. At $r_{1ST} > r_C$, the deepest minimum is no longer on the equator and switches discontinuously to the isolated one close to the north pole. For larger values of $r$, the global minimum approaches the north pole, and the band around the equator shrinks but does not disappear for any finite $r$. The most numerous states, denoted by a dashed line, are always located on the equator.}
    \label{fig:EL3}
\end{figure}

Let us now move back to the more general case in which we consider also a deterministic part $f_D$, which favors configurations in the direction of the signal $\bm{v}$. Without proving them, we now show the results of the Kac-Rice method previously introduced, in three interesting cases.\\
First, let us discuss the case in which $f'_D(0)>0$, which has as simplest example $f_D(x) = x$, corresponding to the spherical $p$-spin model in an external magnetic field of strength $r$. The behaviour with $r$ of the energy landscape in this case is displayed in Fig.~\ref{fig:EL1}. For $r=0$ (the first sphere on the left), there is an exponential number of minima around the equator and the deepest ones are located exactly at the equator, which is also the parallel where the most numerous minima are located. When increasing $r$, the strip containing all the minima moves toward the north pole and starts shrinking, while the deepest minima are on a parallel closer to the north pole as soon as $r > 0$ and thus the most numerous ones are on a different parallel with smaller latitude.
By increasing $r$, the landscape becomes smoother due to a larger deterministic term, and, accordingly, the number of minima and the strip where they are located shrinks until reaching a value $r_C$, called \textit{trivialization point}, above which only one minimum remains and all the other critical points disappear. In this case, the random contribution due to the first term in the Hamiltonian is no longer strong enough to create a rugged landscape but still deforms it sufficiently to move the global minimum at a finite overlap with the signal $\bm{v}$.\\ The second interesting case is the one in which $f_D(x)$  has a vanishing derivative in $x=0$ but a finite second derivative, monotonically increasing from $x=0$ to $x=1$. As a simple example, in Fig.~\ref{fig:EL2} we consider the case $f_D(x) = x^2/2$, which corresponds to a $p$-spin spherical model with an extra ferromagnetic interaction among spins, with r playing the role of the coupling.
The landscape for $r=0$ is the same as before, but by increasing $r$, the strip containing all the minima widens, and the deepest ones and the most numerous ones remain stuck on the equator. This situation persists until $r=r_{2ND}$, at which a \textit{second order phase transition} takes place at the bottom of the landscape; i.e. by increasing $r$ above $r_{2ND}$, the deepest minima continuously detach from the equator. For higher values of $r$, the strip separates into three bands, two closer to the north and south poles, respectively, to which the deepest minima belong, and one around the equator where the most numerous ones are located. At $r=r_C$, the trivialization transition happens, such that the two bands closer to the north and south poles containing an exponential number of minima shrink to zero and are replaced by an isolated global minimum per hemisphere. Finally, at even larger values of $r$, all minima around the equator disappear, and a final transition toward a fully smooth landscape characterized by only two minima takes place.\\ Finally, the last case we discuss is the one in which $f'_D(0) = f''_D(0) = 0$, and the simplest example of such a function is $f_D(x) = x^k$ with $k\geq3$. In Fig.~\ref{fig:EL3}, we focus on this case taking $k=p=3$, corresponding to the spiked-tensor model \cite{arous2019landscape,Lesieur_2017}.
The particularity of this case is that the critical points on the equator are not affected at all by the deterministic perturbation: they remain stable and unperturbed for any finite value of $r$. Therefore, there is always a strip of minima around the equator. One can see that, starting from the same landscape at $r=0$, a band of minima, growing with $r$, is found around the equator. At a value $r_C$, an isolated minimum detaches from the top of the band, and for larger values of $r$ it moves to higher latitudes, while the rest of the band shrinks around the equator. The deepest minima are located on the equator and are the ones of the original (unperturbed) $p$-spin model until a value of $r$, that we call $r_{1ST}$, is reached. When $r$ reaches this value, the global minimum switches from the equator to the single minimum outside the band and close to the north pole. Increasing $r$ further, the isolated global minimum approaches the north pole, and the band around the equator shrinks but never disappears for any finite $r$.
\subsubsection{Further works and applications}
The Kac-Rice method has a vast range of applicability. It has been used not only in physics, but also for example in ecology \cite{ben2021counting,ros2023generalized} to study the number of equilibria in large complex system. Recent reviews can be found in \cite{auffinger2022optimization,ros2022high}, while the quenched computation using Kac-Rice plus replica is in \cite{ros2019complex}. Regarding neural networks, one of the first application of the Kac-Rice method was to compute the number of fixed points for random recurrent neural networks \cite{wainrib2013topological}. Recently, in \cite{pmlr-v107-maillard20a} a teacher-student network was considered, with a loss
\begin{equation}
    \mathcal{L}_{T-S}(\bm{w})=\sum_{\mu}\ell\left(\bm{\xi}_\mu^\top\bm{w^*},\bm{\xi}_\mu^\top\bm{w}\right),
\end{equation}
where $\xi_{\mu}^i \sim \mathcal{N}(0,1)$. This was a methodological advance since the the loss is non-Gaussian. The Kac-Rice method was generalized to analyse this case.
\section{Dynamical mean-field theory for the perceptron model}
\label{sec:DMFT_perceptron}
In this lecture, we introduce dynamical mean-field theory (DMFT). This technique has a long history in statistical physics, where it has been used to analyse the high-dimensional dynamics of strongly correlated disordered systems \cite{SZ81,SZ82,CK93}. DMFT has also been widely employed in condensed matter theory to describe strongly correlated electrons \cite{MV89,GKKR96} resulting in significant advances in both theory and applications.  

This method has a great potential to study the dynamics of high-dimensional problem in machine learning.   
We introduce it in a simple setting, corresponding to the perceptron problem. 
We focus on gradient-flow dynamics in a prototypical learning problem, namely the teacher-student perceptron in dimension $N$. The training dataset $\mathcal{D}=\{(\bm{\xi}_\mu, y_\mu)\}_{\mu=1}^M$ of size $M=\alpha N$, $\alpha\sim\mathcal{O}(1)$, is made of $N-$dimensional i.i.d.~Gaussian samples $\xi_\mu^i\sim\mathcal{N}(0,1/N)$, $\forall i=1,\ldots,N$, $\forall \mu=1,\ldots,M$, and teacher-generated labels: $y_\mu=\phi(\bm{\xi}_\mu^\top\bm{w^*})$. The teacher vector $\bm{w^*}$ is drawn uniformly at random on the hypersphere of radius $\sqrt{N}$: $\bm{w^*}\in\mathcal{S}^{N-1}(\sqrt{N})$. The optimization is performed via gradient descent on the empirical risk:
\begin{align}
    \mathcal{L}(\bm{w})=\sum_{\mu=1}^M\ell\left(\bm{\xi}_\mu^\top\bm{w^*},\bm{\xi}_\mu^\top\bm{w}\right),
\end{align}
where the weight vector $\bm{w}$ is constrained on the hypersphere $\bm{w}\in\mathcal{S}^{N-1}(\sqrt{N})$ at each step of the dynamics. For simplicity, we have incorporated the dependence on the activation function $\phi$ in the loss function $\ell$. This formulation encompasses different widely-studied settings, some celebrated examples being
\emph{binary teacher-student classification}: $\phi(\cdot)={\rm sign}(\cdot)$ \cite{gardner1989three},
and the \emph{sign retrieval problem}: $\phi(\cdot)=(\cdot)^2$, i.e., \emph{phase retrieval} in real space (see \cite{dong2022phase} and references therein).

The initial condition is drawn at random $\bm{w}(0)\sim P_{0}$, where the initial distribution $P_0$ does not depend on the dataset. The gradient flow dynamics is defined as follows:
\begin{align}
\frac{\dd}{\dd t}{\bm{w}}(t)=-\grad_{\bm{w}}\mathcal{L}(\bm{w}(t))-\lambda(t)\bm{w}(t)
\end{align}
where $\lambda(t)$ is a Lagrange multiplier enforcing the spherical constraint $\sum_{i=1}^N w_i(t)^2=N$ at all training times. We aim at describing the system in the infinite-dimensional (a.k.a. \emph{thermodynamic}) limit $N\rightarrow\infty$. To this end, we look for low dimensional order parameters evolving according to a self-consistent equation that effectively characterizes the dynamics of the high-dimensional system.
\subsection{The dynamical cavity method} In order to achieve this effective characterization, we employ the \emph{dynamical cavity method} \cite{MPV87,agoritsas2018out,liu2021dynamics}. In particular, we follow the derivation introduced in \cite{liu2021dynamics} in the context of high-dimensional interacting particle systems. We start by identifying the three reference directions that are relevant in our problem:
\begin{itemize}
    \item The ``lab frame'', a fixed direction $\bm{v}_i$ uncorrelated with the examples $\{\bm{\xi}_\mu\}_{\mu=1}^M$, such that $\bm{w}^\top\bm{v}_i:=w_i$, $\Vert \bm{v}_i\Vert=1$;
    \item The special directions given by the examples $\bm{\xi}_\mu$, that enter in the loss,  such that $\bm{w}^\top\bm{\xi}_\mu:=w_\mu$, where $\Vert \bm{\xi}_\mu\Vert\rightarrow 1$ in the infinite dimensional limit;
    \item The signal (or teacher) $\bm{w}^*$: we call the teacher-student alignment  $m:=\frac{\bm{w}^\top\bm{w^*}}{\sqrt{N}\Vert\bm{w^*}\Vert}$ the \emph{magnetization}, in line with the physics terminology.
\end{itemize}
We now proceed to write a simple self-consistent stochastic process along these key directions. For simplicity, we focus on the random case, where no teacher is present and hence there is no correlation between the random data and labels. At the end, we will discuss how the result modifies if the teacher is introduced and the related generalization properties. In the random case, the loss is only a function of the scalar product $\bm{\xi}_\mu^\top \bm{w}$. We start by writing the dynamics along the first (lab-frame) direction:
\begin{align}
    \frac{\dd}{\dd t}{w}_i(t)=-\sum_{\mu=1}^M\ell'\left(\bm{\xi}_\mu^\top\bm{w}(t)\right)\xi_\mu^i-\lambda(t)w_i(t), \qquad\text{where}\qquad\xi_\mu^i=\bm{\xi}_\mu^\top\bm{v}_i.\label{eq:dynamics_vi}
\end{align}
The remaining degrees of freedom $\bm{w}_{\perp}=\bm{w}-w_i\bm{v}_i$, orthogonal to $\bm{v}_i$, follow the dynamics
\begin{align}
    \frac{\dd}{\dd t}{\bm{w}}_\perp(t)=-\sum_{\mu=1}^M\ell'\left(\bm{\xi}_\mu^\top\bm{w}(t)\right)\bm{\xi}_\mu^\perp-\lambda(t)\bm{w}_\perp(t), \qquad\text{where}\qquad \bm{\xi}_\mu^\perp=\bm{\xi}_\mu-\xi_\mu^i\bm{v}_i.
\end{align}
 Similarly as when deriving the Langevin equation from Newton equations, we can solve the equations for the remaining degrees of freedom $\bm{w}_\perp$ (akin to the ``environment'') at fixed $w_i$, and then plug this solution in Eq. \eqref{eq:dynamics_vi} in order to get a closed equation on $w_i$.\\
 We remind that $\xi_\mu^i\sim\mathcal{N}(0,1/N)$, therefore at large $N$, by the central limit theorem: $\xi_\mu^iw_i~\sim~\mathcal{O}(1/\sqrt{N})$ and $\bm{w}_\perp^\top\bm{\xi_\mu}^\perp\sim\mathcal{O}(1)$. It follows that $\bm{\xi}_\mu^\top\bm{w}=\bm{w}_\perp^\top\bm{\xi_\mu}^\perp+\mathcal{O}(1/\sqrt{N})$ and we can solve for $\bm{w}^\perp$ by using perturbation theory up to linear order. The zeroth-order term in perturbation theory reads
\begin{align} 
\frac{\dd}{\dd t}\bm{w}_\perp^0=-\sum_{\mu=1}^M\ell'\left((\bm{\xi}_\mu^\perp)^\top \bm{w}_\perp^0(t)\right)\bm{\xi}_\mu^\perp-\lambda(t)\bm{w}_\perp^0(t)\label{eq:zeroth_order}
 \end{align}
 with random initial condition on $\bm{w}_\perp^0\in\mathbb{R}^{N-1}$. Note that Eq.~\eqref{eq:zeroth_order} has a well-defined solution for $\bm{w}_\perp^0$. We can now compute the solution for $\bm{w}_\perp$ as a linear order perturbation to $\bm{w}_\perp^0$. To this end, we add the infinitesimal field $h_\mu(t)=\xi_\mu^iw_i(t)$ to the argument of the loss function: $\ell\left((\bm{\xi}_\mu^\perp)^\top \bm{w}_\perp^0(t)\right)\leftarrow \ell\left((\bm{\xi}_\mu^\perp)^\top \bm{w}_\perp^0(t)+h_\mu(t)\right)$. We find
 \begin{align}
     \bm{w}_\perp(t)=\bm{w}_\perp^0(t)+\sum_{\mu=1}^M\int_0^t\dd t' \,\frac{\delta \bm{w}^0_\perp(t)}{\delta h_\mu(t')}\biggr\rvert_{h_\mu=0}\xi_\mu^iw_i(t'),\label{eq:solution_perturb}
 \end{align}
 and we neglect higher order terms in the perturbation. At this point, we can plug the solution for $\bm{w}_\perp$ given by Eq.~\eqref{eq:solution_perturb} into Eq.~\eqref{eq:dynamics_vi} to obtain a closed equation for the dynamics of $w_i$:
\begin{align}
    \frac{\dd }{\dd t}w_i(t)=-\sum_{\mu=1}^M\ell'\left((\bm{\xi}_\mu^\perp)^\top\bm{w}_\perp(t)+\xi_\mu^iw_i(t)\right)\xi_\mu^i-\lambda(t)w_i(t)\\
    =\underbrace{-\sum_{\mu=1}^M\ell'\left((\bm{\xi}_\mu^\perp)^\top\bm{w}^0_\perp(t)\right)\xi_\mu^i}_{\rm (I):\; Random\; force}\;\underbrace{-\sum_{\mu,\mu'=1}^M\int_0^t\dd t'\frac{\delta \ell'\left((\bm{\xi}_\mu^\perp)^\top\bm{w}^0_\perp(t)\right)}{\delta h_{\mu'}(t')}\biggr\rvert_{h_{\mu'}=0}\xi_\mu^i\xi_{\mu'}^iw_i(t')}_{\rm (II):\;Retarded \;friction}\\
    \underbrace{-\sum_{\mu=1}^M\ell''\left((\bm{\xi}_\mu^\perp)^\top\bm{w}_\perp^0(t)\right)(\xi_\mu^i)^2w_i(t)}_{\rm (III):\;Dynamic\; renormalization\;of\;regularization}\;-\lambda(t)w_i(t),\label{eq:dynamics_wi}
\end{align}
that is correct up to linear order in the perturbation. We now briefly comment on the physical meaning of the different terms appearing in Eq.~\eqref{eq:dynamics_wi}. 
\begin{list}{}{}
    \item[(I)] The zeroth-order term in the perturbation is usually called \emph{random force} in physics:
    $$F_i(t)=-\sum_{\mu=1}^M\ell'\left((\bm{\xi}_\mu^\perp)^\top\bm{w}_\perp^0(t)\right)\xi_\mu^i.$$
    Notice that $(\bm{\xi}_\mu^\perp)^\top\bm{w}_\perp^0$ is uncorrelated from $\xi_\mu^i$, therefore $F_i(t)$ is a Gaussian function in the thermodynamic limit, with zero mean $\left<F_i(t)\right>=0$ and covariance 
    \begin{align*}
    \left<F_i(t)F_i(t')\right>&=\left<\sum_{\mu=1}^M(\xi_\mu^i)^2\ell'\left((\bm{\xi}_\mu^\perp)^\top\bm{w}^0_\perp(t)\right)\ell'\left((\bm{\xi}_\mu^\perp)^\top\bm{w}^0_\perp(t')\right)\right>_{\bm{w}_\perp,\bm{\xi}_\mu}\\
    &=\frac 1N\sum_{\mu=1}^M\left<\ell'\left((\bm{\xi}_\mu^\perp)^\top\bm{w}^0_\perp(t)\right)\ell'\left((\bm{\xi}_\mu^\perp)^\top\bm{w}^0_\perp(t')\right)\right>_{\bm{w}_\perp,\bm{\xi}_\mu^\perp}\\
    &=\alpha\left<\ell'\left(\bm{\xi}_\mu^\top\bm{w}(t)\right)\ell'\left(\bm{\xi}_\mu^\top\bm{w}(t')\right)\right>_{\bm{w},\bm{\xi}_\mu}:=M(t,t').
    \end{align*}
    The last equality is obtained by observing that all the examples denoted by $\mu\in\{1,\ldots,M\}$ are statistically equivalent, and putting back in the loss argument the $\mathcal{O}(1/\sqrt{N})$ correction that is negligible in the average.
    \item[(II)] The second term contains the correction to linear order in the perturbation. The contribution coming from $\mu=\mu'$ concentrates in the high-dimensional limit:
    \begin{align*}
        \qquad -\sum_{\mu=1}^M\int_0^t\dd t'\frac{\delta\ell'\left((\bm{\xi}_\mu^\perp)^\top\bm{w}_\perp^0(t)\right)}{\delta h_{\mu}(t')}\biggr\rvert_{h_\mu=0}(\xi_\mu^i)^2\, w_i(t')\\\overset{N\rightarrow\infty}{\longrightarrow}\;-\,\alpha\int_0^t\dd t' \left<\frac{\delta\ell'\left(\bm{\xi}_\mu^\top\bm{w}(t)\right)}{\delta h_{\mu}(t')}\biggr\rvert_{h_\mu=0}\right>_{\bm{w},\bm{\xi}_\mu}w_i(t'):=-\,\alpha\int_0^t\dd t'\,R(t,t')\,w_i(t'),
    \end{align*}
 where we have reintroduced the perturbation in the loss argument without changing the result, similarly as before. We have denoted the above average by $R(t,t')$ since it is a response function. Notice that the terms in $\mu\neq\mu'$ vanish. Overall the term (II) plays the role of a \emph{dissipation} or \emph{retarded friction}. This contribution is also called \emph{Onsager reaction term}.
 \item[(III)] The third term concentrates in a similar way
 \begin{align*}
     -\sum_{\mu=1}^M\ell''\left((\bm{\xi}_\mu^\perp)^\top\bm{w}_\perp^0(t)\right)(\xi_\mu^i)^2w_i(t)\;\overset{N\rightarrow\infty}{\longrightarrow}\;-\alpha\left<\ell''\left(\bm{\xi}_\mu^\top\bm{w}(t)\right)\right>_{\bm{w},\bm{\xi}_\mu}w_i(t):=-\alpha\nu(t)w_i(t),
 \end{align*}
 and results in a dynamical renormalization of the regularization term. 
\end{list}
Finally, we can regroup all the terms above and write the stochastic equation for the dynamics of $w_i$:
\begin{align}
    \frac{\dd }{\dd t}w_i(t)=-\alpha\int_0^t\dd t'\,R(t,t')w_i(t')+F_i(t)-\left(\lambda(t)+\alpha\nu(t)\right)w_i(t),\label{eq:final_wi}
\end{align}
with random Gaussian initial condition $w_i(0)\sim\mathcal{N}(0,1)$. The noise comes from the Gaussian force $F_i(t)$, with zero mean and covariance $\left<F_i(t)F_i(t')\right>=M(t,t')$.\\
At this point, we need one last ingredient to close the equations. Indeed, the argument of the loss function depends on the projection of the weigths $\bm{w}$ onto the direction of the examples $\bm{\xi}_\mu$. Therefore, we need to derive a dynamical equation for this order parameter. The dynamics along the direction $\bm{\xi}_\mu$ is given by
\begin{align}
    \frac{\dd}{\dd t}w_\mu&=-\sum_{\mu'=1}^M\ell'\left(\bm{\xi}_{\mu'}^\top\bm{w}(t)\right)\bm{\xi}_\mu^\top\bm{\xi}_{\mu'}-\lambda(t)w_\mu(t)\\
    &=-\sum_{\mu'(\neq\mu)}\ell'\left(\bm{\xi}_{\mu'}^\top\bm{w}(t)\right)\bm{\xi}_\mu^\top\bm{\xi}_{\mu'}-\lambda(t)w_\mu(t)-\ell'\left(\bm{\xi}_{\mu}^\top\bm{w}(t)\right)\underbrace{\Vert\bm{\xi}_\mu\Vert^2}_{=1}.\label{eq:eq_wu}
\end{align}
The above Eq.~\eqref{eq:eq_wu} is now formally identical to the one for $w_i$ (Eq.~\eqref{eq:dynamics_vi}), with only two differences: the sum in this second case runs over $M-1$ examples ($\mu\neq\mu'$) and there is an additional term $-\ell'\left(\bm{\xi}_{\mu}^\top\bm{w}(t)\right)$. The first difference is negligible in the high-dimensional limit. Moreover, the examples $\bm{\xi}_{\mu'}$ with $\mu'\neq\mu$ are uncorrelated from $\bm{\xi}_\mu$. Therefore, if there was no extra term in Eq.~\eqref{eq:eq_wu}, the effective stochastic process for the dynamics of $w_\mu$ would be the same as in Eq.~\eqref{eq:final_wi}. Including the extra term, we obtain
\begin{align}
    \frac{\dd }{\dd t}w_\mu(t)=-\alpha\int_0^t\dd t'\,R(t,t')w_\mu(t')+F_\mu(t)-\left(\lambda(t)+\alpha\nu(t)\right)w_\mu(t)-\ell'\left(w_\mu(t)\right),\label{eq:final_wu}
\end{align}
with Gaussian initial condition on  $w_\mu(0)$ and the same definitions as in Eq.~\eqref{eq:final_wi} for $R(t,t')$, $M(t,t')=\langle F_\mu(t)F_\mu(t')\rangle$ and $\nu(t)$:
\begin{align}
    R(t,t')&:=\frac{\delta\left<\ell'(w_\mu(t))\right>_{w_\mu}}{\delta h_\mu(t')}\biggr\rvert_{h_\mu=0},\\
    M(t,t')&:=\alpha\left<\ell'(w_\mu(t))\ell'(w_\mu(t'))\right>_{w_\mu},\\
    \nu(t)&:=\left<\ell''(w_\mu(t))\right>_{w_\mu}.
\end{align}
Finally, we need to specify an equation to compute the Lagrange multiplier $\lambda(t)$ enforcing the spherical constraint $\sum_{i=1}^N w_i(t)^2=N$, $\forall t\geq0$. By taking the derivative with respect to $t$ on both sides and using Eq.~\eqref{eq:dynamics_vi}, we find 
\begin{align}
    \sum_{i=1}^Nw_i(t)\frac{\dd}{\dd t}w_i(t)=-\sum_{\mu=1}^M\ell'\left(\bm{\xi}_\mu^\top\bm{w}(t)\right)\bm{\xi}_\mu^\top \bm{w}(t)-\lambda(t)\underbrace{\sum_{i=1}^Nw_i(t)^2}_{=N}=0.\label{eq:lagrange_multi}
\end{align}
Dividing both sides of the above Eq.~\eqref{eq:lagrange_multi} by $N$, we obtain that in the high-dimensional limit
\begin{align}
    \lambda(t)=-\alpha\left<\ell'(w_\mu)w_\mu\right>_{w_\mu}.
\end{align}
It is important to remark that the dynamical mean-field equations derived above are expressed in terms of a self-consistent stochastic process. Indeed, the equations depend on the kernels $R$ and $M$ and the auxiliary function $\nu$, that are in turn obtained as averages over the same stochastic process. By causality, it can be shown that the solution of the DMFT system is unique. This circular structure highlights the ``mean-field'' nature of these equations, similarly as the celebrated equation for the magnetization: $m={\rm th}(\beta m)$ for the Ising model at inverse temperature $\beta$. 

Interestingly, the above equations -- derived here with an heuristic method -- have been put on rigorous ground in some cases \cite{arous1997symmetric,arous2001aging,ben2006cugliandolo,celentano2021high,gerbelot2022rigorous}.

Armed with the effective description of the high-dimensional gradient-descent dynamics provided by DMFT, we can now proceed to analyze the case where the dataset includes labels generated by a teacher vector $\bm{w^*}$, representing a prototype supervised learning problem. As previously anticipated, this  modification introduces another important order parameter, i.e., the \emph{teacher-student overlap} or \emph{magnetization} $m(t)=\bm{w}(t)^\top\bm{w^*}$. One way to obtain an effective equation for $m(t)$ is to notice that $\sum_{i=1}^N{w_i}(t)^\top{w^*_i}/N\overset{N\rightarrow\infty}{\longrightarrow}\left<w_i^*w_i(t)\right>_{w_i,w_i^*}$. By multiplying Eq.~\eqref{eq:final_wi} by $w_i^*$ and taking the average, we find the following ODE:
\begin{align}
    \frac{\dd }{\dd t}m(t)=-\alpha\int_0^t\dd t'\,R(t,t')m(t')-\left(\lambda(t)+\alpha\nu(t)\right)m(t).
\end{align}
The above equations allow us to study the learning curves of the problem, for instance:
\begin{itemize}
    \item The dynamical evolution of the average loss function: $$\frac{1}{M}\mathcal{L}(\bm{w}(t))=\frac 1M\sum_{\mu=1}^M\ell\left(\bm{\xi}_\mu^\top\bm{w^*};\bm{\xi}_\mu^\top\bm{w}(t)\right)\overset{N\rightarrow\infty}{\longrightarrow}\left<\ell(w^*_\mu,w_\mu(t))\right>_{w_\mu};$$
    \item The dynamical evolution of the magnetization,\footnote{In this spherical case, the generalization error: $\mathbb{E}\left[\mathbf{1}\left(\phi(\bm{\xi}^\top\bm{w})\neq y\right)\right]$, with $\mathbf{1}(\cdot)$ denoting the indicator function, is monotonic decreasing in the magnetization. Therefore, the magnetization captures all the relevant information on the performance.} that we can use to investigate signal recovery and its time scales;
    \item The properties of the correlation function: $C(t,t')=\sum_{i=1}^Nw_i(t)w_i(t')/N\overset{N\rightarrow\infty}{\longrightarrow}\left<w_i(t)w_i(t')\right>$.
\end{itemize}

It is important to remark that we have started with a \emph{deterministic} dynamics in high dimensions, and we have ended up with an effective dynamics involving a \emph{random} force and a dissipation term. This is the result of isolating a representative variable while integrating out all the remaining degrees of freedom, as it happens when studying the dynamics of many physical systems (e.g., a molecule in a liquid, a spin in a magnetic material).

Another important point to underline is that the infinite-dimensional limit $N\rightarrow\infty$ is taken at \emph{fixed} time window $[0,t]$. Therefore, this method cannot address timescales that diverge with the system size $N$.

\subsubsection{Special case: the spherical spin glass}
It is instructive to mention a special case where the DMFT equations considerably simplify. This is the spherical $p-$spin glass model \cite{crisanti1992sphericalp}, described by the disordered long-range $p-$body Hamiltonian
\begin{align}
    H(\bm{s})=\;-\sum_{i_1,i_2,\ldots,i_p}J_{i_1,i_2,\ldots,i_p}s_{i_1}s_{i_2}\ldots s_{i_p}\,, \quad {\rm with }\;\;\;\sum_{i=1}^N s_i^2=N,
\end{align}
where we have denoted the degrees of freedom by $\bm{s}\in\mathcal{S}^{N-1}(\sqrt{N})$ for historical reasons. The symmetric tensor $J_{i_1,i_2,\ldots,i_p}$ of rank $p$ can be either drawn from a standard i.i.d.~Gaussian distribution or generated by a teacher vector.
We can implement the same procedure as above and obtain an effective equation for the $i^{\rm th}$ spin:
\begin{align}
\frac{\dd }{\dd t}s_i(t)=\underbrace{F(t)}_{\rm random \;force}+\frac{p(p-1)}{2}\int_0^t\dd t'\,R(t,t')\,C^{p-2}(t,t')s_i(t')-\lambda(t)s_i(t),
\end{align}
where
\begin{align}
    \langle 
    F(t)F(t')
    \rangle&:=C(t,t')=\langle s_i(t)s_i(t')\rangle,\\
    R(t,t')&:=\frac{\delta\langle s_i(t)\rangle}{\delta h_i(t')}\biggr\rvert_{h_i=0}.
\end{align}
The above equations are much simpler than the perceptron case since the effective stochastic process is a linear equation with additive noise and can be solved to write closed equations on the correlation and response functions $C$ and $R$.  
\\\\
In summary, DMFT equations are a very powerful dimensional reduction tool to describe the dynamics of disordered systems at fixed time windows. Interesting extensions of the solution derived here are possible, for instance studying discrete variables with a Monte Carlo approach \cite{eissfeller1994mean}, Langevin noise\footnote{Notice that in the presence of Langevin noise in the dynamics particular care must be taken when computing derivatives, as in Eq.~\eqref{eq:lagrange_multi} where Itô's rule must be used.} \cite{mannelli2020marvels,agoritsas2018out}, momentum-based accelerated methods \cite{sarao2021analytical}, stochastic gradient descent \cite{mignacco2020dynamical}.

A current bottleneck and important future development of the method regards the numerical implementation of the solution. The simplest strategy is to discretize the DMFT system, start by an initial guess for the kernels/auxiliary functions, use this guess to generate multiple realizations of the stochastic process, compute the averages and iteratively update the kernels $M_{\rm new}(t,t')=(1-\gamma)M_{\rm old}(t,t')+\gamma M_{\rm new}(t,t')$ until convergence, where $\gamma$ is an appropriately chosen damping factor. This procedure was first implemented in the context of theoretical ecology \cite{roy2019numerical}. A key challenge for future research is to improve these numerical solvers, similarly as what has already been done in the case of quantum problems, leading to great theoretical and application advances.


\section{Energy barriers, entropic barriers, and signal recovery}
In the following, we want to spell out some important features of gradient dynamics in high-dimensional non-convex landscapes. In the next paragraphs, we analyze the gradient flow algorithm for a simple toy model of signal recovery problems. First, we discuss how the gradient flow can be trapped by spurious local minima in a complex landscape, causing failure of signal recovery. Next, we consider a stochastic version of gradient flow dynamics and discuss the so-called "entropic barriers" coming from the dynamics' randomness as another cause of failure in recovery. At the end of this lecture, we discuss the connection with dynamics in deep neural networks.

\subsection{Good and bad minima for gradient flow dynamics in high-dimensions}

In this section, we discuss whether the gradient flow (GF) dynamics on non-convex energy landscapes can recover a signal hidden within a random noise following the works \cite{mannelli2020marvels,sarao2019afraid,mannelli2019passed}. Within this setting we discuss the phenomenon of "bad and good minima" that is attracting a lot of attention in machine learning \cite{liu2020bad}. 

\begin{figure}[ht]
\centering
 \includegraphics[width=0.8\linewidth]{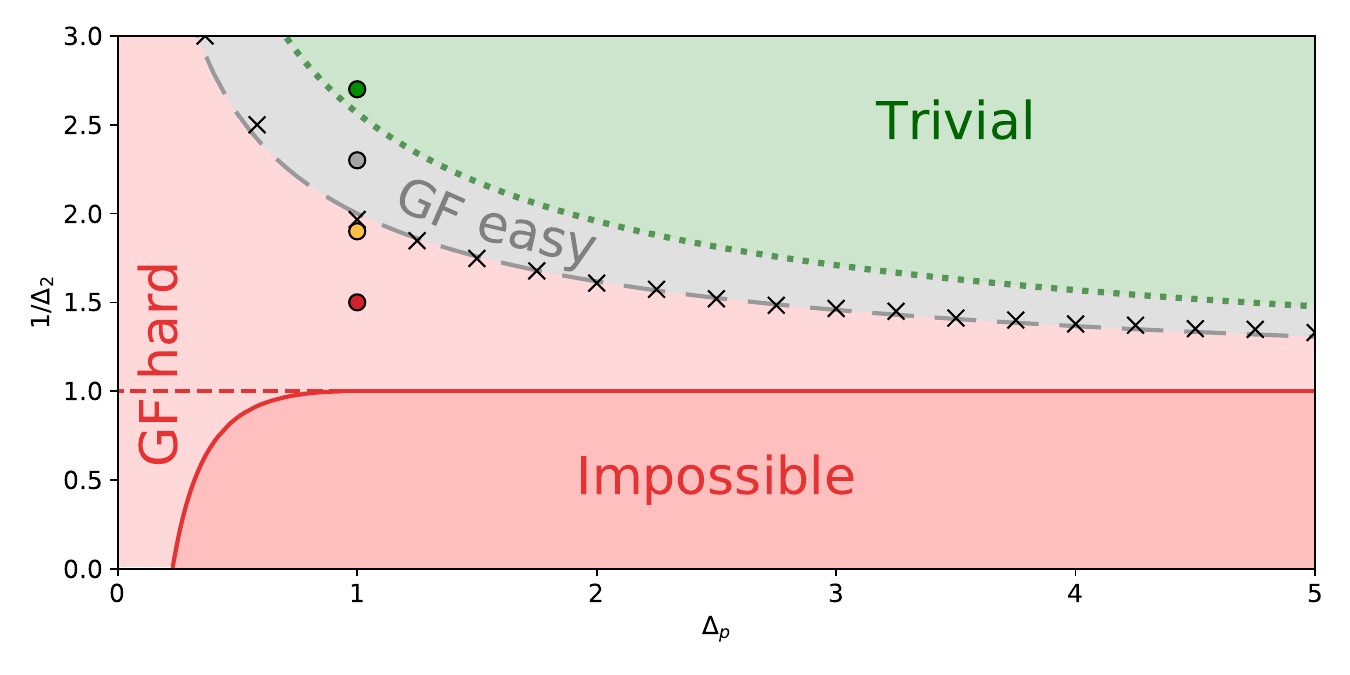}
 \caption{Phase diagram of the gradient flow for the spiked Matrix Tensor model. In the dark red region, it is information-theoretically impossible to obtain correlation with the signal\cite{mannelli2020marvels}. In the light red region, while there exist minima correlated to the signal, the gradient flow typically cannot find them. In the green region, the energy landscape does not have spurious local minima. The gradient descent works well in the grey region between these two regions despite many spurious local minima. This figure is taken from S. Sarao Mannelli {\it et al.}, 2019 \cite{sarao2019afraid}.}
 \label{fig:phase-diagram}
\end{figure}

The model we consider here is the so-called spiked Matrix-Tensor model. Our task is to retrieve a signal on ($N-1$)-dimensional sphere $\sigma^*\in\mathbf{S}^{N-1}=\{v\in \mathbb{R}^N: \|v\| = 1\}$ by observing a matrix $\mathbf{Y}$ and a tensor $\mathbf{T}$ given by
\begin{eqnarray}
T_{i_{1} \ldots i_{p}}&=&\eta_{i_{1} \ldots i_{p}}+\sqrt{N(p-1) !} \,\sigma_{i_{1}}^{*} \ldots \sigma_{i_{p}}^{*} \nonumber\\
Y_{i j}&=&\eta_{i j}+\sqrt{N} \,\sigma_{i}^{*} \sigma_{j}^{*},
\end{eqnarray}
where $\eta_{i_{1} \ldots i_{p}}$ and $\eta_{i j}$ are independent centered Gaussian random variables with variance $\Delta_p$ and $\Delta_2$ respectively. We here assume that $p>2$. The maximum likelihood estimation of the signal $\sigma^*$ corresponds to the minimization of an energy function $\mathcal{H} = \mathcal{H}_s + \mathcal{H}_p$, where the signal part $\mathcal{H}_s$ and the noisy part $\mathcal{H}_p$ are given by
\begin{eqnarray}
    \mathcal{H}_s &=& -\frac{1}{\Delta_2 \sqrt{N}} \sum_{i<j} Y_{ij}\sigma_i\sigma_j\nonumber\\
     \mathcal{H}_p &=& -\frac{\sqrt{(p-1)!}}{\Delta_p \sqrt{N}} \sum_{i_1<i_2<\cdots<i_p} T_{i_1i_2\cdots i_p}\sigma_{i_1}\cdots \sigma_{i_p}.
\end{eqnarray}
This minimization is performed by GF on the sphere $\mathbf{S}^{N-1}$, initializing from a uniform random point on the sphere. To understand the performance of GF, we can apply the dynamical mean-field theory and the Kac-Rice formula \cite{sarao2019afraid}. These analyses result in the phase diagram Fig.\ref{fig:phase-diagram}. In the bottom part of this phase diagram, it is impossible to recover the signal due to the absence of minima associated with the signal. In the left part of the figure with $1/\Delta_2<1.0$, there is a hard phase, where there exists a minimum with a finite correlation with the signal, but it is difficult to find it by the Approximate Message Passing algorithm (a very good algorithm for this problem). While the Approximate Message Passing algorithm succeeds in finding the signal above $1/\Delta_2 >1.0$, GF fails below the dashed line. Interestingly, this line is well below the landscape trivialization threshold (the dotted line) computed by the Kac-Rice formula. This shows a fact that can be astonishing at first sight: GF can find the minimum associated with the signal while the energy landscape still has exponentially many spurious minima (grey-colored region). The mechanism behind this phenomenon can be understood by dynamical mean-field theory and the Kac-Rice method, as we shall explain below.

\begin{figure}[ht]
 \begin{minipage}{0.4\hsize}
  \begin{center}
   \includegraphics[width=0.9\linewidth]{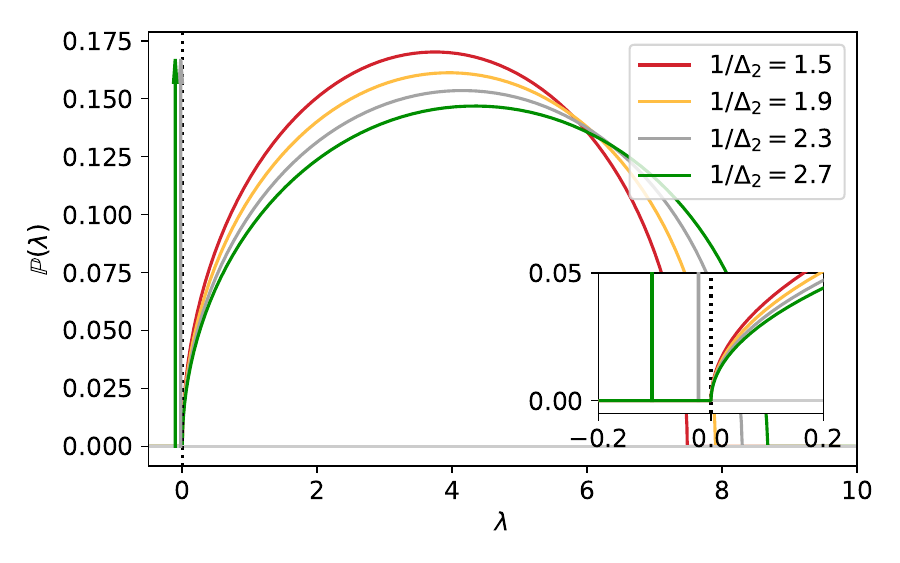}
  \end{center}
 \end{minipage}
 \begin{minipage}{0.29\hsize}
  \begin{center}
   \includegraphics[width=0.9\linewidth]{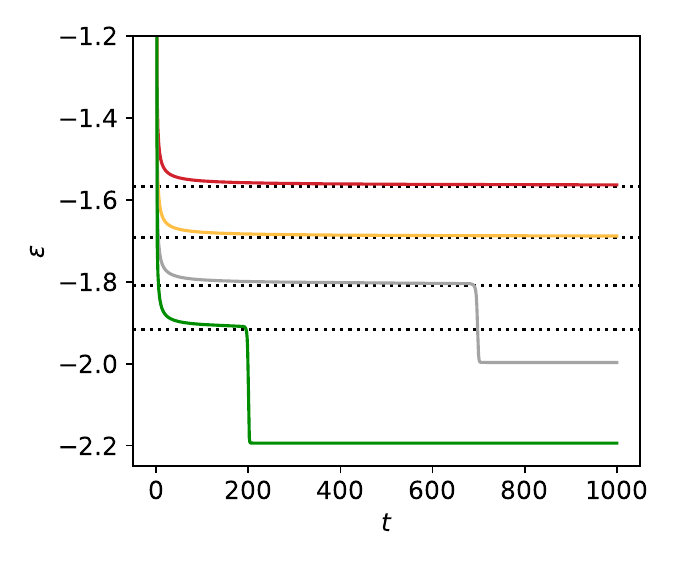}
  \end{center}
 \end{minipage}
 \begin{minipage}{0.29\hsize}
  \begin{center}
   \includegraphics[width=0.9\linewidth]{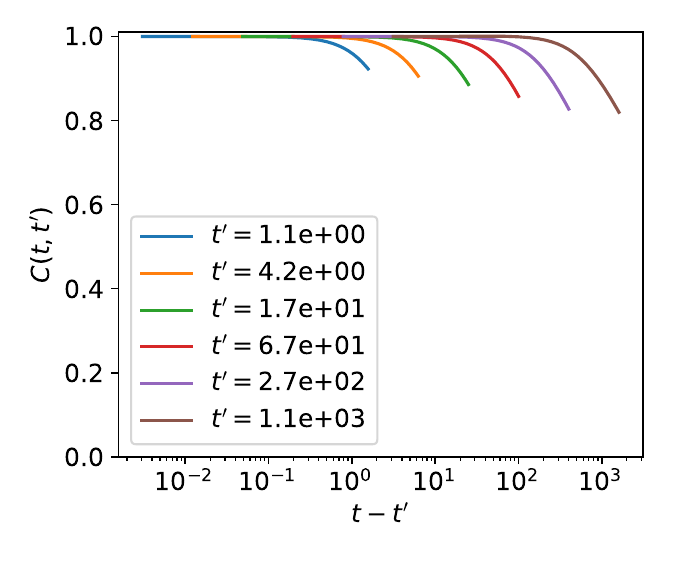}
  \end{center}
 \end{minipage}
 \caption{
 Left Panel: The hessian spectrum of threshold states with parameter values corresponding to the four points of the same color in Fig.\ref{fig:phase-diagram}. When $\Delta_2$ is large enough, a single isolated eigenvalue appears depicted as an arrow, destabilizing the threshold states. This destabilization allows the gradient flow dynamics suddenly go further down in the energy landscape, as we can observe in the center panel.\\
 Center panel: The energy trajectories of the gradient flow dynamics. Each trajectory corresponds to the point on the phase diagram Fig.\ref{fig:phase-diagram} of the same color. The dotted line is the energy of threshold states computed by dynamical mean-field analysis and the Kac-Rice analysis. \\
 Right panel: The correlation function $C(t, t')$ with $p = 3, \Delta_p = 1.0$ and $1/\Delta_2 = 1.5$ numerically obtained from
the dynamical mean-field analysis.
 These figures are taken from S. Sarao Mannelli {\it et al.}, 2019 \cite{sarao2019afraid}.}
 \label{fig:2+3_hessian_dynamics_energy}
\end{figure}

In the region of the phase diagram where GF fails, the numerical integration of the dynamical mean-field equation shows that GF relaxes into a certain energy level, well above the lowest energy (the red and yellow curves in the center panel of Fig.\ref{fig:2+3_hessian_dynamics_energy}). As the left panel of Fig.\ref{fig:2+3_hessian_dynamics_energy} shows, the auto-correlation function $C(t,t')$ decreases as $t-t'$ increases but its relaxation time scale goes larger as the age of the system $t'$ increases. This phenomenon is called aging in the physics literature \cite{CK93}. We will discuss more in detail later. In this aging regime, we observe that the state after a long time $t\to\infty$ is marginally stable, i.e., the eigen-spectrum of its Hessian is a shifted-semicircle whose left edge touches zero (the left panel of Fig.\ref{fig:2+3_hessian_dynamics_energy}). The Kac-Rice analysis reveals that the asymptotic energy (the dotted lines in the center panel of Fig.\ref{fig:2+3_hessian_dynamics_energy}) is the threshold energy where the left edge of the Hessian spectrum of typical critical points touches zero. These minima in this energy level are the most numerous and very flat; therefore, we can expect that they have large basins of attraction, which is why GF tends to converge to those points in this regime.
When $\Delta_2$ is large enough, similarly, the energy trajectory first converges to the threshold energy. Interestingly, however, it then suddenly drops within a finite time (the grey and green curves in the center panel of Fig.\ref{fig:2+3_hessian_dynamics_energy}), and eventually the state reaches a minimum with a finite overlap with the signal. This successful escape from the threshold states is due to their BBP transition, which can be seen by the Kac-Rice analysis. 
As the left panel of Fig.\ref{fig:2+3_hessian_dynamics_energy} shows, when the signal-to-noise ratio $\Delta_2$ is large enough, an isolated eigenvalue pops out from the bulk of Hessian's eigen-spectrum of the threshold states, and the corresponding eigenvector has a finite overlap with the signal $\sigma^*$. This isolated eigenvalue is strictly negative on the threshold energy, and hence the threshold states are typically unstable. This unstable direction has a finite overlap with the signal. By following it the system approaches the (good) minimum correlated with the signal. These two dynamical regimes (until the threshold and then toward the signal) are analogous to the search and convergence phase already discussed in Sec. 3. 

\begin{figure}[ht]
\centering
 \includegraphics[width=0.6\linewidth]{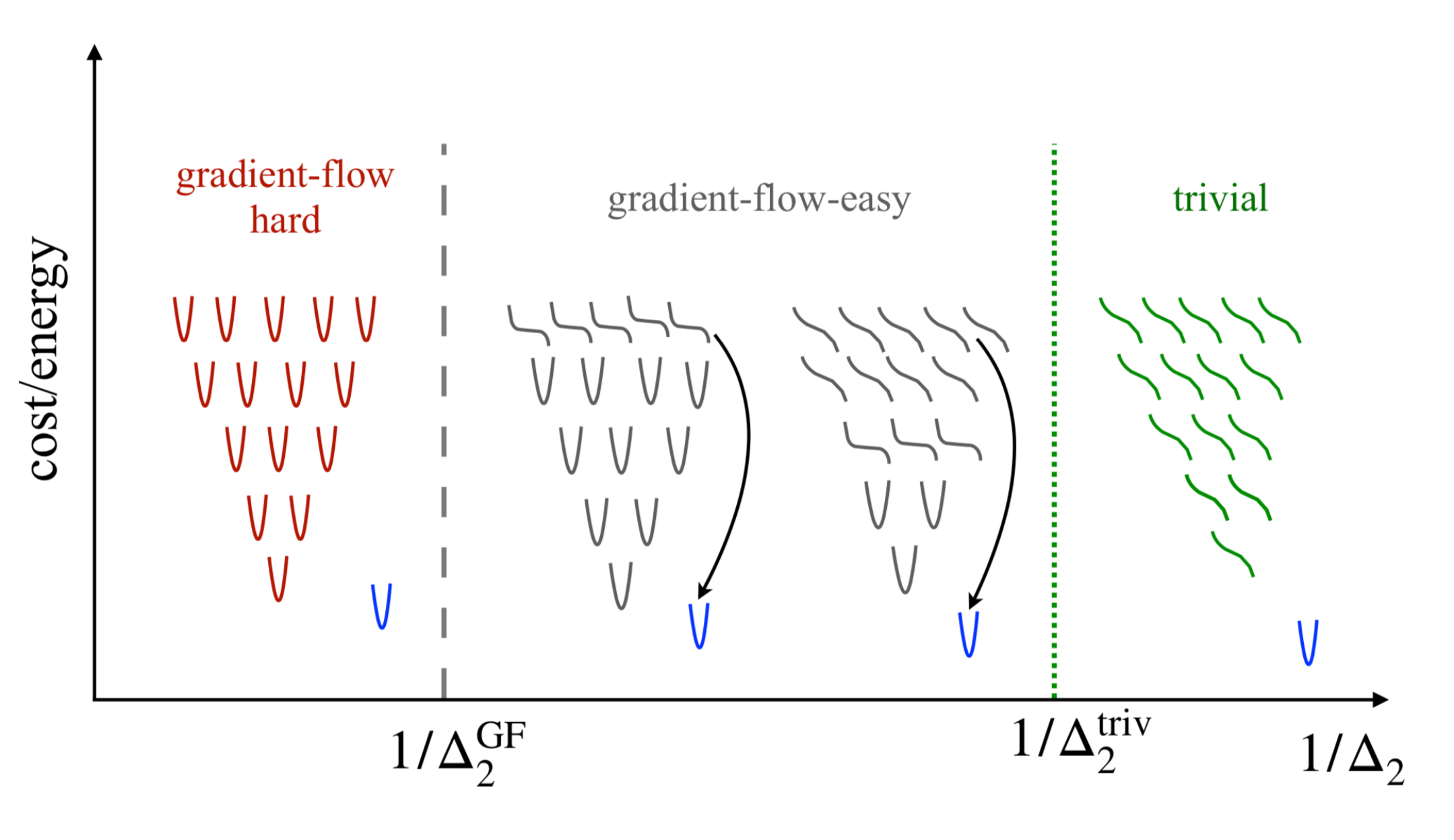}
 \caption{Cartoon of the phase transitions of the energy landscape of the spiked Matrix-Tensor model. This figure is taken from S. Sarao Mannelli {\it et al.}, 2019 \cite{sarao2019afraid}.}
 \label{fig:landscape-cartoon}
\end{figure}

Fig.\ref{fig:landscape-cartoon} summarizes the phase transitions as we increase the SNR. If the SNR is small, GF is typically trapped by the threshold states, which have large basins of attraction but no correlation with the signal. As we increase the SNR, however, the threshold states become unstable due to the BBP transition, and the unstable direction navigates the dynamics toward a minimum correlated to the signal. Once the SNR gets large enough, all the other spurious minima become unstable, and thus the energy landscape becomes trivial.

\subsection{Entropic barriers}
This section discusses another mechanism hampering gradient descent dynamics: the so-called entropic barriers \cite{unpub,arous2020algorithmic, arous2022high,bellitti2021entropic}. We consider a very simple setting to discuss the main mechanism at play \cite{unpub}: the online Stochastic Gradient dynamics with the Tensor-PCA model. At each time step, the Gaussian random tensor is independently chosen, i.e., the covariance between the tensor $\mathbf{J}^t$ sampled at time $t$ and $\mathbf{J}^{t'}$ sampled at time $t'$ is given as follows
\begin{equation}
    \langle J^t_{i_1,i_2,\cdots,j_p} J^{t'}_{i_1,i_2,\cdots,j_p}\rangle = \frac{(p-1)!}{N^{p-1}}\delta(t-t').
\end{equation}
The gradient is given by
\begin{equation}
    \frac{\partial E(\sigma|J^{t})}{\partial \sigma_i} =  -\sum_{i_1<i_2<\cdots<i_{p-1}} \left(J^t_{ii_1\cdots i_{p-1}} + J^t_{i_1i\cdots i_{p-1}} +\cdots + J^t_{i_1\cdots i_{p-1}i}\right) \sigma_{i_1}\cdots \sigma_{i_{p-1}} - r\left(\sum_j \sigma_j\sigma^*_j\right)^{p-1} \sigma^*_i,
\end{equation}
where $r$ here is the signal-to-noise ratio.
For simplicity, we here analyze the gradient flow, a continuous limit of discretized gradient descent.
Let $\bar\xi_i(t)$ denote the first term (without the negative sign) on the right-hand side. Then, the time derivative of $\sigma_i$ is obtained as
\begin{equation}
    \frac{d\sigma_i}{dt} = - \frac{\partial E(\sigma|J^{t})}{\partial \sigma_i} = rm^{p-1} \sigma^*_i+ \bar \xi_i(t) -\lambda(t) \sigma_i,
    \label{eq:ds_idt}
\end{equation}
where $m := N^{-1}\sum_j \sigma_j\sigma^*_j$, and $\lambda(t)$ is a Lagrange multiplier to constrain $\mathbf{\sigma}$ on the sphere $\mathbf{S}^{N=1}$.

It is easy to see that $\bar \xi(t)$ is a Gaussian noise with zero mean $\langle\bar \xi(t)\rangle = 0$. The variance $T:=\langle\bar \xi_i^2(t)\rangle$ has the order of $O(1)$, as is shown by the following calculation.
\begin{eqnarray}
    \langle\bar \xi_i(t)\bar\xi_i(t')\rangle
    &=&  \sum_{\substack{i_1<i_2<\cdots<i_{p-1}\\j_1<j_2<\cdots<j_{p-1}}} \sigma_{i_1}\cdots \sigma_{i_{p-1}} \sigma_{j_1}\cdots \sigma_{j_{p-1}}\nonumber\\
    &&\qquad\times\langle\left(J^t_{ii_1\cdots i_{p-1}} +\cdots + J^t_{i_1\cdots i_{p-1}i}\right) \left(J^{t'}_{ij_1\cdots j_{p-1}} + \cdots + J^{t'}_{j_1\cdots j_{p-1}i}\right)\rangle \nonumber\\
    &=& \frac{p!}{N^{p-1}}\sum_{\substack{i_1<i_2<\cdots<i_{p-1}}}\sigma_{i_1}^2\cdots \sigma_{i_{p-1}}^2 \delta(t-t')\nonumber\\
    &=& p\delta(t-t'),
\end{eqnarray}
where we have neglected sub-leading terms in $N$ (due to the large $N$ limit).

\begin{figure}[ht]
\centering
 \includegraphics[width=0.5\linewidth]{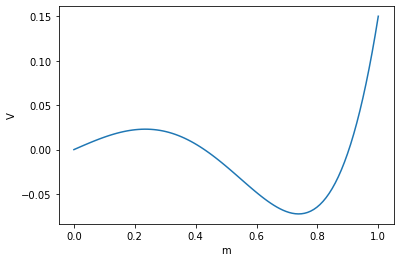}
 \caption{The effective potential function $V(m) = Tm - r(1-m^2)m^{p-1}$ shows the entropic barrier. The parameter values are chosen as follows: $T=0.2, r= 1.0, p = 4$.}
 \label{fig:potential-m}
\end{figure}

The time derivative of $m$ can be obtained from Eq.\eqref{eq:ds_idt} by multiplying $\sigma^*_i/N$ and taking the summation over index $i$, that is
\begin{equation}
    \frac{dm}{dt} = rm^{p-1} - \lambda(t) m + A(t).
    \label{eq:dm-dt-1}
 \end{equation}
Here $A(t) := N^{-1}\sum_i \sigma^*_i \bar \xi_i$ is of the order of $O(\frac{1}{\sqrt{N}})$. Next, we identify the Lagrange multiplier $\lambda$ by imposing the condition of $\sum_i \sigma_i^2=N$, which means $\frac{d}{dt}\sum_i \sigma_i^2= 0$. Exploiting the Ito's formula,
\begin{eqnarray}
    \frac{1}{2N}\frac{d}{dt} \sum_i \sigma_i^2 &=& \frac{1}{2N}\sum_i\frac{d \sigma_i^2}{d\sigma_i} \frac{d\sigma_i}{dt} +\frac{T}{2N} \sum_i\frac{d^2\ \sigma_i^2}{d\sigma_i^2}\nonumber\\
    &=& \frac{1}{N}\sum_i\sigma_i \frac{d\sigma_i}{dt} +T\nonumber\\
    &=& rm^{p} -\lambda(t) + B(t) + T,
\end{eqnarray}
where $B(t) := N^{-1}\sum_i \sigma_i \bar \xi_i = O(1/\sqrt{N})$ is the stochastic part. Hence,
\begin{equation}
    \lambda(t) = rm^p + B(t) + T.
\end{equation}
Substituting this to Eq.\eqref{eq:dm-dt-1},
\begin{eqnarray}
    \frac{dm}{dt} &=& - (T + rm^p) + rm^{p-1} + A(t) + B(t)\nonumber\\
    &=& -\frac{\partial V}{\partial m} + \mathrm{ noise}.
\end{eqnarray}
where the effective potential is $V = Tm - r(1-m^2)m^{p-1}$, drawn in Fig.\ref{fig:potential-m}, and the noise is very weak (of order $1/\sqrt{N})$. When $r$ is large enough, one finds two minima in the potential; one is at the origin, the other at large $m$ and there is a barrier of order one separating them. Since $m(0) = O(1/\sqrt{N})$ with the random initialization, the dynamics start around the local minimum at the origin. Since the barrier height is much larger than the noise amplitude, one needs to wait an exponentially long time in $N$ to climb up the barrier to find the global minimum at large $m$, i.e. the signal cannot be recovered in polynomial time in $N$. In consequence, the noise in the equation on $m$ is completely ineffective and we can neglect it to understand the dynamical behavior. 
The barrier in the potential is due to the first term proportional to the noise amplitude $T$. In fact, because of the noise,  the system is kicked randomly at each time step by the drawn sample of the tensor, which typically brings the state back to the equator (i.e., the region of $m\approx 0$) since the measure on the sphere concentrates around its equator. Hence, this barrier is due to the large entropy of configurations around the equator, which we call an "Entropic Barrier". Note that when $r$ is large enough $r/T \gg N^{(p-2)/2}$, the width of the entropic barrier is much smaller than $1/\sqrt{N}$, and therefore the dynamics start from the right side of the entropic barrier, in which case the global minimum is reachable.

The main conclusion of this simple analysis is that noise is not always beneficial. In low dimensions it helps navigating the landscape and escaping bad minima, however in high dimensions it can bring the system in high-entropy configurations not correlated with the signal, whereas in other cases it can help if these regions have good generalization properties \cite{annesi2023star}. 

\subsection{Numerical experiments in deep neural networks}
\begin{figure}[ht]
 \begin{minipage}{0.5\hsize}
  \begin{center}
   \includegraphics[width=0.9\linewidth]{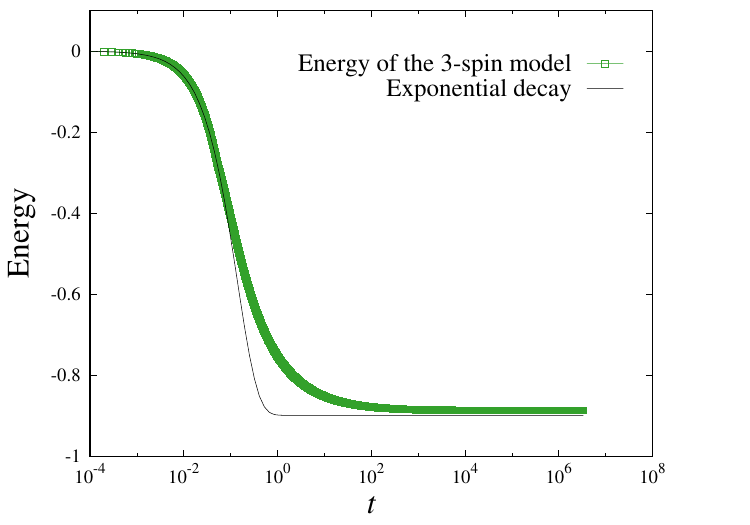}
  \end{center}
 \end{minipage}
 \begin{minipage}{0.5\hsize}
  \begin{center}
   \includegraphics[width=0.9\linewidth]{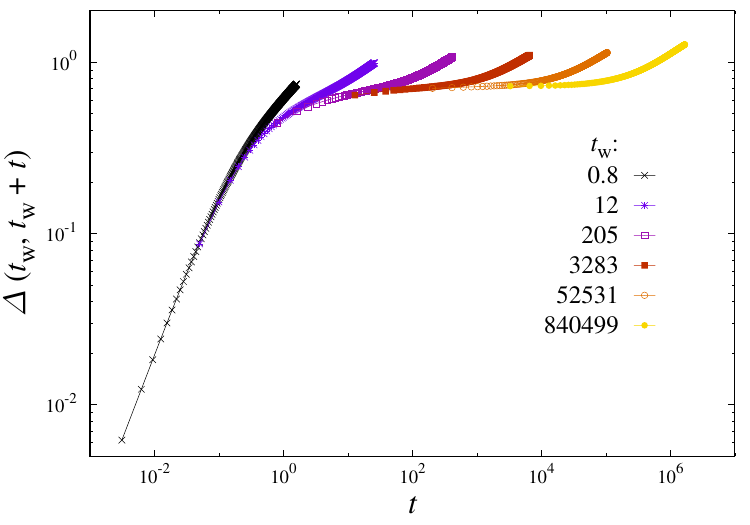}
   
  \end{center}
 \end{minipage}
 \caption{The trajectories of quenched dynamics of the 3-spin model from a high temperature $T_i=\infty$ to a low temperature $T_f = 0.5$. In the left panel, we plot the energy trajectory in comparison with an exponentially decaying curve. In the right panel, the mean square displacement is displayed for several values of $t_w$. These figures are taken from M. Baity-Jesi {\it et al}, 2018 \cite{baity2018comparing}.}
 \label{fig:pspin-dynamics}
\end{figure}

In the previous sections, we analyzed simple theoretical models with techniques from glass physics. In this section, we discuss the empirical observation of deep neural networks comparing it with glassy aging dynamics, based on M. Baity-Jesi {\it et al.}, 2018  \cite{baity2018comparing}. The aim of this work was to investigate whether the training dynamics is glassy or to what extent out of equilibrium using tools and observables developed in statistical physics. To this aim, we first briefly revisit the glassy dynamics of the spherical 3-spin model -- an archetypical model of glasses. Its energy reads
\begin{equation}
    E=-\sum_{\left\langle i_{1}, i_{2}, i_{3}\right\rangle} J_{i_{1}, i_{2}, i_{3}} \sigma_{i_{1}} \sigma_{i_{2}} \sigma_{i_{3}}.
\end{equation}
Here the summation goes over all the possible triplets of indexes running from $1$ to $N$, and the coupling $J_{i_{1}, i_{2}, i_{3}}$ are i.i.d. centered Gaussian random variables with variance $3/N^2$. The spin configuration $\sigma$ is a $N$-dimensional vector on the sphere of radius $\sqrt{N}$. The plots in Fig.\ref{fig:pspin-dynamics} correspond to stochastic Langevin dynamics under a quench from high temperature $T_i=\infty$ to low temperature $T_f = 0.5$. As we can observe in the left panel, the relaxation to the asymptotic energy is slower than exponential decay, which is a characterization of the aging phenomenon in glassy systems. Another characteristic of aging can be observed in the mean-square displacement, defined as
\begin{equation}
    \Delta\left(t_{\mathrm{w}}, t_{\mathrm{w}}+t\right)=\frac{1}{N} \sum_{i=1}^{N}\left(\sigma_{i}\left(t_{\mathrm{w}}\right)-\sigma_{i}\left(t_{\mathrm{w}}+t\right)\right)^{2}.
\end{equation}
The right panel of Fig.\ref{fig:pspin-dynamics} shows the trajectories of the mean-square displacement against $t$ with various fixed values of $t_w$. It clearly shows that as the age of the system $t_w$ gets larger, it takes more time to decorrelate the system. This is another attribute of the aging phenomenon. Note that these phenomena are quite general, and are displayed by many physical glassy systems \cite{berthier2011theoretical}.

\begin{figure}[ht]
 \begin{minipage}{0.5\hsize}
  \begin{center}
   \includegraphics[width=0.8\linewidth]{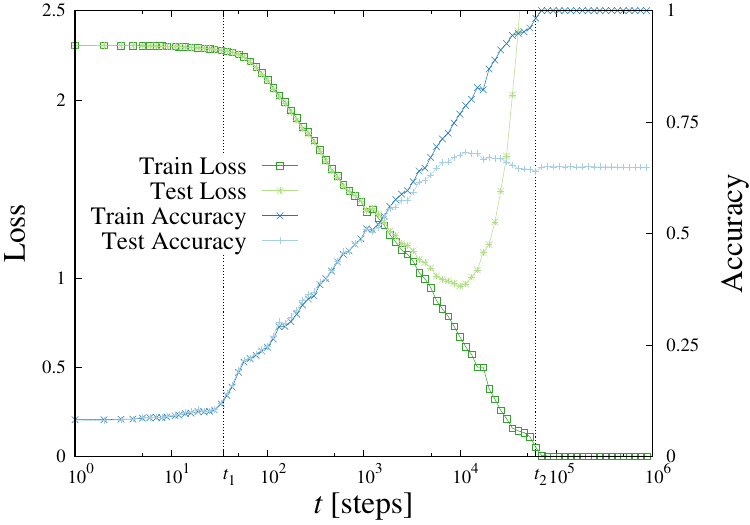}
  \end{center}
 \end{minipage}
 \begin{minipage}{0.5\hsize}
  \begin{center}
   \includegraphics[width=0.8\linewidth]{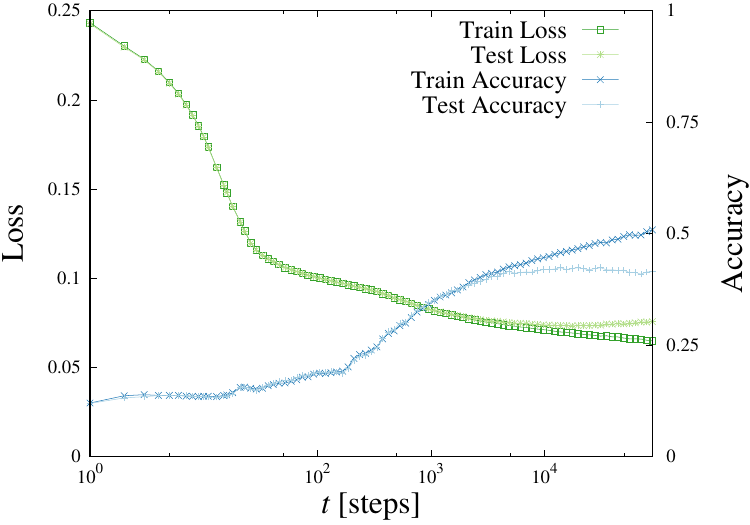}
  \end{center}
 \end{minipage}
 \caption{Train/test loss and accuracy of neural networks trained for CIFAR-10 as a function of $\log(t)$. The left and right panel displays the models with over-parameterization and under-parameterization, respectively. These figures are taken from M. Baity-Jesi {\it et al.}, 2018  \cite{baity2018comparing}.}
 \label{fig:loss-dnn}
\end{figure}

Now we discuss the learning dynamics of the deep neural networks comparing it with the glassy dynamics. Fig.\ref{fig:loss-dnn} shows the trajectories of loss as a function of time steps when we train neural networks for CIFAR-10. In the left panel, the model is over-parameterized. This case shows three regimes during the training process. At first, it explores in the high-loss configurations up to $t = t_1$. Next, the loss decreases approximately linearly in $\log(t)$,
and the accuracy increases similarly. This regime is up to time $t=t_2$, where the training loss goes to zero. In the last regime, the training loss always stays around zero. This behavior resembles the one discussed in the previous sections and consisting in a search and a convergence phase. 
For comparison, we show the under-parameterized case in the right panel of Fig.\ref{fig:loss-dnn}, which shows different behavior from the one with over-parameterization. In this case, the dynamics resembles the one of glassy landscapes in which the system converges to bad minima and non-zero training loss.  

\begin{figure}[ht]
 \begin{minipage}{0.5\hsize}
  \begin{center}
   \includegraphics[width=0.8\linewidth]{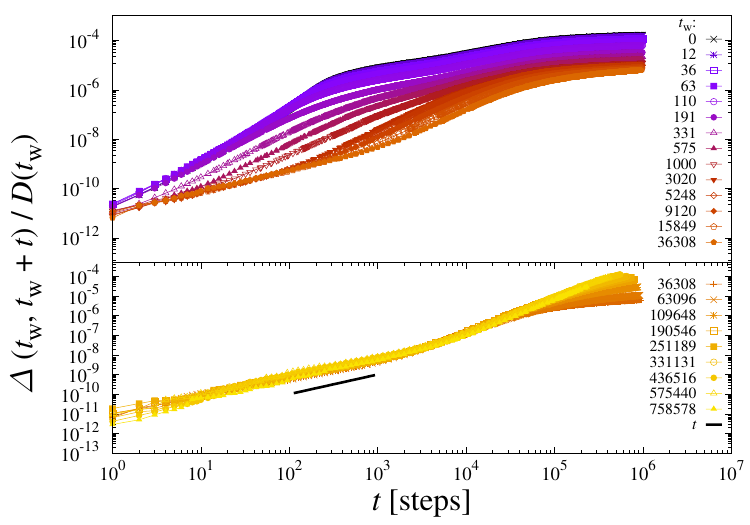}
  \end{center}
 \end{minipage}
 \begin{minipage}{0.5\hsize}
  \begin{center}
   \includegraphics[width=0.8\linewidth]{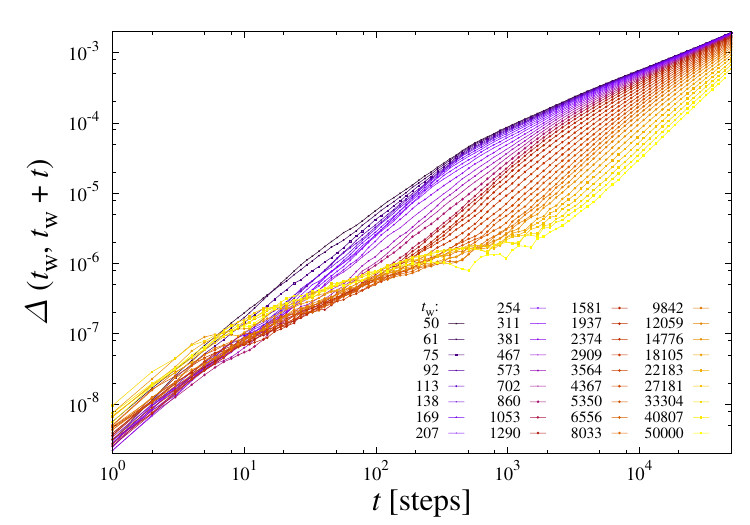}
  \end{center}
 \end{minipage}
 \caption{Mean-squared displacement of the learning trajectories of neural networks as a function of $\log(t)$. The left panel displays an over-parameterized model trained for MNIST, and the right panel displays an under-parameterized model trained for CIFAR-10. These figures are taken from M. Baity-Jesi {\it et al.}, 2018   \cite{baity2018comparing}.}
 \label{fig:msd-dnn}
\end{figure}

Next, we discuss the mean-squared displacement of the neural networks' learning dynamics, shown in Fig.\ref{fig:msd-dnn}. In the left panel, we show the case of the over-parameterized networks trained for MNIST. The three different colors roughly correspond to the three regimes we discussed above. The intriguing observation here is that at the final stage of the dynamics (the yellow curves), up to the re-scaling of the noise amplitude $D(t_w)$, the mean-squared displacement is almost always independent of the age of the system $t_w$. On the other hand, the under-parameterized regime shown in the right panel of Fig.\ref{fig:msd-dnn} shows the aging phenomenon similar to glassy systems. 

The training dynamics of over-parameterized neural networks displays interesting phenomena. In the under-parametrized regime, one finds aging dynamics and slow convergence to bad minima, whereas in the over-parametrized one the dynamics it has a search phase and a convergence phase. At long times, it becomes stationary if one renormalizes the unit of time, corresponding to diffusion over the zero (or very small) training loss manifold. The theoretical understanding of the transitions between aging and non-aging dynamics and the three regimes during the training process has still to be completed, as shown here insights from physics can be helpful on this endeavour. 
Filling the gap between the theoretical toy models we understand so far and the empirical observations in the learning dynamics of neural networks is an important open problem.

\section*{Acknowledgments}
We thank F. Krzakala and L. Zdeborov\'a for organizing the summer school "Statistical Physics and Machine Learning". 

\paragraph{Author contributions}
These lecture notes are based on a series of lectures give by G. Biroli at the Les Houches Summer School "Statistical Physics and Machine Learning". TB, DG, KK, FM, AY contributed equally by preparing these lecture notes. GB revised them. 

\paragraph{Funding information}
GB acknowledge funding from the French government under management of Agence Nationale de la Recherche as part of the ``Investissements d’avenir'' program, reference ANR-19-P3IA-0001 (PRAIRIE 3IA Institute) and from the Simons Foundation collaboration ``Cracking the Glass Problem'' (No. 454935 to G. Biroli). 
FM and KK were supported in part by the National Science Foundation, through the Center for the Physics of Biological Function (PHY-1734030). KK was also supported by a C.V. Starr Fellowship.


\newpage
\bibliography{refs.bib}
\bibliographystyle{unstr}

\end{document}